\documentclass[proof]{pasj00}
\begin{document}
\SetRunningHead{RXTE Observations of MAXI J1659--152}{Running Head}

\title{Combined Spectral and Timing Analysis of the Black Hole Candidate MAXI J1659--152 Discovered by MAXI and Swift}

\author{Kazutaka \textsc{Yamaoka}\altaffilmark{1},
Ryan \textsc{Allured}\altaffilmark{2},
Philip \textsc{Kaaret}\altaffilmark{2},
Jamie A. \textsc{Kennea}\altaffilmark{3},
Toshihiro \textsc{Kawaguchi}\altaffilmark{4},
Poshak \textsc{Gandhi}\altaffilmark{5},
Nicholai \textsc{Shaposhnikov}\altaffilmark{6},\altaffilmark{7},
Yoshihiro \textsc{Ueda}\altaffilmark{8}, 
Satoshi \textsc{Nakahira}\altaffilmark{9},
Taro \textsc{Kotani}\altaffilmark{10},
Hitoshi \textsc{Negoro}\altaffilmark{11},
Ichiro \textsc{Takahashi}\altaffilmark{1}, 
Atsumasa \textsc{Yoshida}\altaffilmark{1}, 
Nobuyuki \textsc{Kawai}\altaffilmark{12}, and 
Satoshi \textsc{Sugita}\altaffilmark{13}}
\altaffiltext{1}{Department of Physics and Mathematics, Aoyama Gakuin University, 5-10-1 Fuchinobe, Sagamihara, Chuo-ku, Kanagawa 252-5258, Japan}
\email{yamaoka@phys.aoyama.ac.jp}
\altaffiltext{2}{Department of Physics and Astronomy, University of Iowa, Van Allen Hall, Iowa City, IA 52242, USA}
\altaffiltext{3}{Department of Astronomy and Astrophysics, Pennsylvania State University, University Park, PA 16802, USA}
\altaffiltext{4}{Center for Computational Sciences, University of Tsukuba, Ten-nodai, 1-1-1 Tsukuba, Ibaraki 305-8577, Japan}
\altaffiltext{5}{Institute of Space and Astronautical Science (ISAS), Japan Aerospace Exploration Agency (JAXA), 3-1-1 Yoshino-dai, Chuo-ku, Sagamihara, Kanagawa 252-5210, Japan}
\altaffiltext{6}{Department of Astronomy, CRESST/University of Maryland, College Park, MD 20742, USA}
\altaffiltext{7}{Goddard Space Flight Center, NASA, Astrophysics Science Division, Greenbelt MD 20771, USA}
\altaffiltext{8}{Department of Astronomy, Kyoto University, Oiwake-cho, Sakyo-ku, Kyoto 606-8502, Japan}
\altaffiltext{9}{MAXI team, RIKEN, 2-1 Hirosawa, Wako, Saitama 351-0198, Japan}
\altaffiltext{10}{Research Institute for Science and Engineering, Waseda University, 17 Kikui-cho, Shinjuku-ku, Tokyo 162-0044, Japan}
\altaffiltext{11}{Department of Physics, Nihon University, 1-8-14 Kanda-Surugadai, Chiyoda-ku, Tokyo 101-8308, Japan}
\altaffiltext{12}{Department of Physics, Tokyo Institute of Technology, 2-12-1 Ookayama, Meguro-ku, Tokyo 152-8551, Japan}
\altaffiltext{13}{EcoTopia Science Institute, Nagoya University, Furo-cho, Chikusa-ku, Nagoya 464-8603, Japan}

%

\maketitle
\begin{abstract}

We report on X-ray spectral and timing results of the new black hole candidate (BHC) 
MAXI J1659--152 with the orbital period of 2.41 hours (shortest among BHCs) in the 
 2010 outburst from 65 Rossi X-ray Timing Explorer (RXTE) observations and 
 8 simultaneous Swift and RXTE observations.  According to the definitions of the spectral states 
 in \citet{review1}, most of the observations have been classified into the intermediate state.
  All the X-ray broadband spectra can be modeled by a multi-color disk
  plus a power-law with an exponential cutoff or a multi-color disk plus a Comptonization component.
 During the initial phase of the outburst, a high energy cutoff was visible at 30--40 keV. 
 The innermost radius of the disk gradually decreased by a factor of more than 3 
 from the onset of the outburst and reached a constant value of 35 $d_{10}{\cos i}^{-1/2}$ km, 
 where $d_{10}$ is the distance in units of 10 kpc and $i$ is the inclination.  
 The type-C quasi-periodic oscillation (QPO) frequency varied from 1.6 Hz to 7.3 Hz 
 in association with a change of the innermost radius, while 
  the innermost radius remained constant during the type-B QPO detections at 1.6--4.1 Hz. 
 Hence, we suggest that the origin of the type-B QPOs is different from that of type-C 
 QPOs, the latter of which would originate from the disk truncation radius. Assuming the 
 constant innermost radius in the latter phase of the outburst as the innermost stable 
 circular orbit, the black hole mass in MAXI J1659--152 is estimated to be 3.6--8.0 
 $M_{\odot}$ for a distance of 5.3--8.6 kpc and an inclination angle of 60--75 degrees.

\end{abstract}

\section{Introduction}

 Black hole candidates (BHCs) are very important objects for the understanding of 
 accretion and jets in high energy astrophysics (see \cite{review1}, \cite{review2}, and \cite{review3} 
 for a review). Most of them are close X-ray binaries consisting of a black hole and 
 a low mass companion (Cyg X-1, LMC X-1 and X-3 are persistent high mass exceptions)
 characterized by outbursts separated by quiescent periods. During a typical outburst, 
 they go through the hard state (or low/hard state), steep power-law state (SPL; or very high 
 state), intermediate state (IMS), and thermal state (or high/soft state), then the hard state 
 and back to the quiescence, although some other state classifications are 
 present (e.g. soft-intermediate and hard-intermediate state in \cite{review3}).
 
 In the hard state, the X-ray spectrum can be approximately explained by a power-law with an 
 exponential cutoff at 100--200 keV. The power spectrum is characterized by 
  strong variability (20--30 \%) with band-limited noise. 
 The hard X-ray emission likely originates from thermal Comptonization 
 in a high temperature corona ($\sim$10$^9$ K) of 
 soft photons from the disk. On the other hand, the thermal state spectrum is 
 dominated by a soft component which is considered to be a blackbody 
 from the geometrically thin, optically thick accretion disk ($\sim$10$^7$ K). 
 This X-ray spectrum is well fit by the multi-color disk (MCD) 
 model (\cite{mcd1}, \cite{mcd2}), and also by more accurate accretion disk models (e.g. \cite{kerrbb}, 
 \cite{bhspec}) which have been developed to take into account relativistic effects.  
 The derived innermost radius from the MCD model typically remains constant
 at a value which might correspond to the innermost stable circular orbit (ISCO),
 from which an estimate of the BH mass can be found. 
 Thus derived, black hole mass from X-ray data is found to be consistent with the BH mass 
 estimated from the binary kinematics (\cite{constrin_isco}, \cite{mass}). 

Compared to the two main states, both the SPL and IMS are not well understood. 
 They frequently appear during state transitions between the hard and thermal states. 
 Both disk and power-law components are clearly present in the energy spectra, and 
 the power spectra show strong time variability associated with low-frequency QPOs (LFQPOs), 
  seen at 0.1--10 Hz (\cite{review1}). Several types of LFQPOs have been identified 
 and classified into type A, B and C by \citet{qpo_def} and \citet{qpo_def2}. Type A and B QPOs 
 are characterized by a relatively low coherence (the coherence parameter $Q\lesssim$3 for type A 
 and $Q\gtrsim$6 for type B) and a narrow frequency range of 1--8 Hz associated with a 
 weak red-noise component.  Features for the type-C QPOs include a high coherence parameter ($Q\gtrsim$10), 
  a variable frequency (0.1---10 Hz), and strong broadband noise under the QPO. 
 The study of LFQPOs is essential to our understanding of the spectral state transitions in BHCs, 
   but their origin is still a topic of debate. 

 A new hard X-ray transient was first discovered 
 as Gamma Ray Burst (GRB) 100925A on September 25, 2010 (MJD 55464) by the Swift Burst Alert 
 Telescope (BAT; \cite{bat}) \citep{discovery1}. The Gas Slit Camera (GSC; \cite{gsc}) 
 on board the Monitor of All-sky X-ray Image (MAXI; \cite{maxi}) 
 independently detected this X-ray transient and localized it to 
 (RA, Dec)=(16$^{\rm h}$59$^{\rm m}$10$^{\rm s}$, --15$^{\circ}$16'05'') 
 with a 0.2 degree accuracy \citep{discovery2}. 
 The source was designated as MAXI J1659--152. The MAXI/GSC data showed 
 that, unlike normal GRBs, the X-ray flux rapidly increased following the discovery. 
 Rossi X-ray Timing Explorer (RXTE; \cite{rxte}) follow-up observations revealed
  the source as a BHC from spectral and timing properties \citep{rxte_bhc}. 
 Due to the Swift BAT prompt GRB trigger and the source's 
  location well above the Galactic plane, many multi-wavelength observations 
 including radio \citep{radio}, submm bands \citep{submm}, near-infrared \citep{nir}, 
 optical \citep{opt}, X-rays and GeV gamma-rays \citep{GeV} have been carried out 
 since the discovery. The XMM-Newton, RXTE, and Swift data revealed the
 presence of X-ray dips in their light curves with a period of 2.41 hours 
 (\cite{newton_dip}, 2011, \cite{rxte_dip}, \cite{swift_dip}). 
 The amplitude modulation with this period was also confirmed by the optical 
 data in the VSNET-team report \citep{optical_period}. These results indicate that MAXI J1659--152 
 has the shortest orbital period among all the black hole candidates (the second 
 shortest one is 3.2 hours for Swift J1753.5--0127 \citep{swiftj1753}). 

 Several X-ray results have already been reported in \citet{rxte_timing}, \citet{rxte_spectiming} and 
  \citet{rxte_spectiming2}, and \citet{swift_dip}. 
 We shortly summarize results from these four groups who published spectral and 
 timing results using the same RXTE and Swift data sets.
 \citet{rxte_timing} showed a BHC signature based on the hardness-intensity diagram \citep{hid} 
 and timing properties including the LFQPO classification. 
 At almost the same time, \citet{rxte_spectiming} reported on detailed timing properties, including 
 time lags for QPOs.   These two papers mainly focus on timing properties, 
 although they performed spectral fitting with a simple model, i.e. {\tt wabs*(diskbb$+$powerlaw)}. 
 \citet{rxte_spectiming2} performed spectral fits with the 
 bulk-motion Comptonization model ({\tt bmc} in XSPEC; \cite{bmc}), and 
 estimated a BH mass of 20$\pm$3 M$_{\odot}$ and a distance of 7.6$\pm$1.1 kpc 
 using the spectral-timing correlation scaling technique.
 Results from Swift/UVOT, XRT and BAT data covering MJD 55464
  (Day 0) to 55491 (Day 27) were reported by \citet{swift_dip}.  
 They have detected X-ray dips and Type-C QPOs at 0.15--1.9 Hz, particularly during the 
 initial phase of the outburst, and performed broadband fits with 
 XRT and BAT data. 

 In this paper, we report on detailed spectral modeling and combined spectral and 
 timing results using 8 (quasi) simultaneous Swift/XRT and RXTE/PCA observations, as well as  
  the same RXTE data sets. In Sections \ref{obsdata} and \ref{results}, we describe 
 the RXTE follow-up observations and details of spectral and timing 
 analysis of the data. We also show spectral results from Swift/XRT 
 and RXTE/PCA observations.  In Section \ref{discuss}, we discuss how our results compare with 
 those from other BHCs, an estimation of the black hole mass of MAXI J1659--152 from 
 spectral results alone, and the physical origin of the low frequency QPOs. 
 Finally, we will conclude this paper with a summary. 

\section{Observations and Data Analysis}\label{obsdata}

\subsection{RXTE}
 Following the Swift and MAXI discovery of MAXI J1659--152,  
 we triggered RXTE Target of Opportunity (ToO) observations approved in the AO14 Open-Time 
 program to reveal its nature. Our follow-up observations started on 
 September 28, 2010, and two other programs were triggered after our 
 observations and continued until November 8, 2010, when the source was not observable due to the 
 Sun constraint of the satellite. These series of observations contain
 65 pointed observations under the observation IDs 95358, 95108 and 95118. 
The net deadtime-corrected PCA exposure was 134.6 ksec, 
 with an averaged exposure of about 2.1 ksec per pointed observation.

 The RXTE carries three scientific instruments: the Proportional Counter Array 
 (PCA; \cite{pca}), the High Energy X-ray Timing Experiment 
 (HEXTE; \cite{hexte}), and the All-Sky Monitor (ASM; \cite{asm}). The PCA 
 consists of five identical proportional counters (PCU 0--4)
 which are sensitive in the 2--100 keV range, while the HEXTE
 is composed of two identical NaI(Tl)/CsI(Na) phoswich scintillators 
 (we distinguish the detectors as Cluster A and B) which are sensitive in the 15--250 keV 
 range. All three instruments are still active and working well in space since the 
 launch in December 1995. 

 We analyzed the RXTE data using the standard FTOOLS, HEADAS version 6.9,
  provided by NASA Goddard Space Flight Center (GSFC). 
 For the PCA spectral analysis, we used the Standard 2 mode data, which 
 has 16 sec time resolution and 129 energy channels. The selection criteria
 were: (1) the offset angle is larger than 0.02 degrees; (2) the elevation from the Earth 
 is larger than 10 degrees; (3) the PCU2 is active; and (4) there are no PCU breakdowns. 
 We accumulated the energy spectra from only the top layer of the PCU2. Dip events seen in RXTE/PCA and Swift/XRT data were not excluded in our analysis. 
 The PCA background was estimated from the model for bright sources, 
 which is publicly available at the PCA analysis 
 web site\footnote{http://heasarc.gsfc.nasa.gov/docs/xte/pca\_news.html}.
 We performed deadtime corrections (the detector deadtime is typically 
 2--3 \%), and added a 0.5\% systematic error to each energy bin. 
 Based on the improvement to the RXTE/PCA calibration\footnote{Updated on August 26, 2009 at http://www.universe.nasa.gov/xrays/programs/rxte/pca/doc/rmf/pcarmf-11.7/}, the energy range was limited to 3--50 keV.
  XSPEC version 12.5 was 
 used for spectral fits. The errors are quoted at statistical 90\% error. 
 To check the validity of the current response and our analysis method, 
 we analyzed a Crab observation (ObsID: 95802-01-18-00) with a net exposure of
   658 sec, taken on October 22, 2010 during the MAXI J1659--152 outburst. 
  The fit is acceptable ($\chi^2$/dof=62.0/77), and the Crab spectrum in 
  the 3--50 keV range is explained by a power-law with a photon index of 2.12$\pm$0.01 and 
  an observed flux of 2.30$\times$10$^{-8}$ erg cm$^{-2}$ s$^{-1}$ in the 2--10 keV range, which 
  is consistent with the nominal values \citep{crab}.
 
 For the PCA timing analysis, we used the 
 Event mode data\footnote{The Good Xenon mode and Binned mode data were used 
 during some observations.}, which has finer time resolutions than the 
 standard mode data. The power density spectra (PDS) were produced using a light curve 
 binned with a 7.8125 msec time resolution (Nyquist frequency: 64 Hz) and {\tt powspec} in the 
 XRONOS package.  The PDS were normalized to a unit of (rms/mean)$^2$ Hz$^{-1}$ 
 after subtracting the Poisson noise.  

Since January 2010, the pointing directions of HEXTE Cluster A and B have been fixed 
 toward the target position and the 1.5$^{\circ}$-offset background position.  
 Hence, we needed to use the background production tool 
 {\tt hextebackest}, which predicts the background 
 spectrum for Cluster A from that of Cluster B.   
 However, the HEXTE data were not used because the current version of 
{\tt hextebackest} still leaves some systematic features at $\sim$60 keV 
 in the background-subtracted spectra.

\subsection{Swift/XRT}

 BHCs have a typical innermost disk black-body temperature of $\sim$1 keV (i.e. an emission peak of the blackbody at 
 $\sim$2.8 keV) so we only see a Wien tail of the disk blackbody spectrum in the PCA 
 energy range above 3 keV. Hence, there may be large uncertainties in disk parameter estimations of 
 the MCD model, especially when the disk temperature is lower than $\sim$0.6 keV. 
 We, therefore, performed fits of MAXI J1659--152 with Swift X-Ray Telescope (XRT; \cite{swiftxrt}) data, with soft band 
 coverage over 0.3--10 keV, to obtain reliable disk parameters. 
 In the archived XRT data, we found 7 observations that were exactly simultaneous to RXTE observations,
  and one (Period H) that was partially simultaneous.
 The list of simultaneous observations and time intervals
 is shown in Table \ref{tab1}. All the XRT observations reported 
 in this paper were performed in the Windowed Timing (WT) mode. 
  The XRT data were analyzed in the public web interface\footnote{http://www.swift.ac.uk/user\_objects/} \citep{xrt_ana}, and 
  the latest version of the XRT response matrix, swxwt0to2s6\_20070901v012.rmf, was used.  
 The energy range was limited to 0.6--9 keV taking into account
  a calibration uncertainty at low energy and photon statistics at high energy.  

\section{Results}\label{results}
\subsection{PCA Light Curve and State Classification}

\begin{figure}
  \begin{center}
    \FigureFile(80mm,80mm){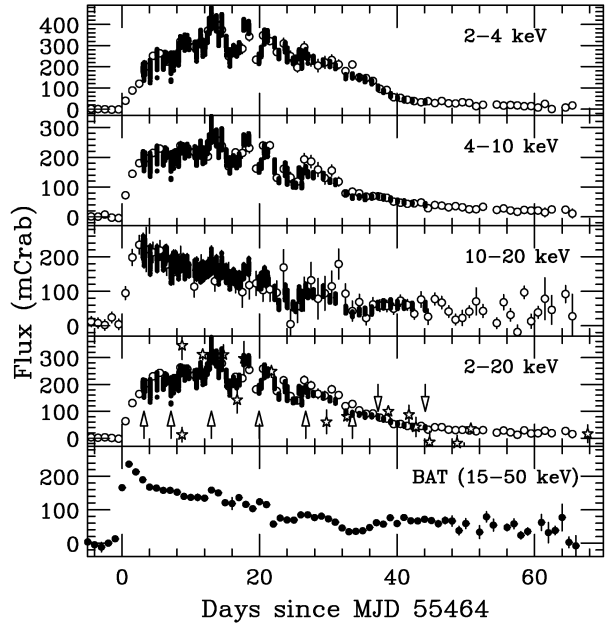}
  \end{center}
  \caption{RXTE/PCA light curves of MAXI J1659--152 during the 2010 outburst. 
 The 2--4, 4--10, 10--20, and 2--20 keV ranges are shown with a 16-sec time bin 
 in the upper four panels. 
 The one-day averaged MAXI/GSC and RXTE/ASM data are superposed on the PCA data 
 with open circles and open stars respectively. 
 The daily averaged Swift/BAT data are also shown in the bottom panel for comparison.
 The arrows in the 4th panel from the top indicate 8 representative observations for which the spectra and the power spectra are plotted in Figure \ref{fig2} and \ref{fig6}, respectively.}\label{fig1}
\end{figure}

The RXTE/PCA PCU2 light curves with a time resolution of 16 sec in four energy ranges (2--4, 4--10, 10--20, 2--20 keV) are shown in Figure \ref{fig1}. The data were 
 normalized to the PCU2 counts for the Crab nebula. The MAXI/GSC\footnote{http://maxi.riken.jp/top}, the RXTE/ASM\footnote{http://heasarc.nasa.gov/docs/xte/ASM/sources.html}, 
 the Swift/BAT\footnote{http://heasarc.nasa.gov/docs/swift/results/transients/} data 
 were taken from the web sites for each instrument team, and also normalized to the Crab and  
 are shown for comparison. 
 The flux measurements between PCA and MAXI are consistent with each other. 
 The soft band flux in the 2--10 keV range slowly rises with some variability,
 while the hard band flux rapidly rises within a few days
  and then slowly decreases. The approximate e-folding time is about 
 30 days at high energies. On October 15, 2010, the peak was about 300 mCrab in the 
 2--20 keV range, and the total outburst duration was about 65 days, 
 assuming the MAXI/GSC and Swift/BAT data cover the whole outburst.
  
 We classified all the observations into accretion states following
 the definitions in \citet{review1}. Note that our classification  is based on fitting 
 results with the {\tt wabs*(diskbb+powerlaw)} model in the 3--50 keV range, 
 with a cutoff and iron line during some of the early observations. We found that all 
 the observations were classified into just two states: 58 IMSs and 7 hard states. 
 Neither the thermal
  state nor the SPL state were observed.  
 The source was initially in the hard state for $\sim$4.1 days, 
 then transitioned into the IMS, where it remained for 32 days. It finally returned to the 
 hard state around MJD 55503 (day 39). No intense 
 power-law flares as seen in , e.g., XTE J1550--564 \citep{xte1550} 
 and H 1743--322 \citep{h1743} were observed. 

\subsection{Spectral fitting results}
\subsubsection{RXTE/PCA}\label{pcafit}

\begin{figure*}
  \begin{center}
    \FigureFile(140mm,140mm){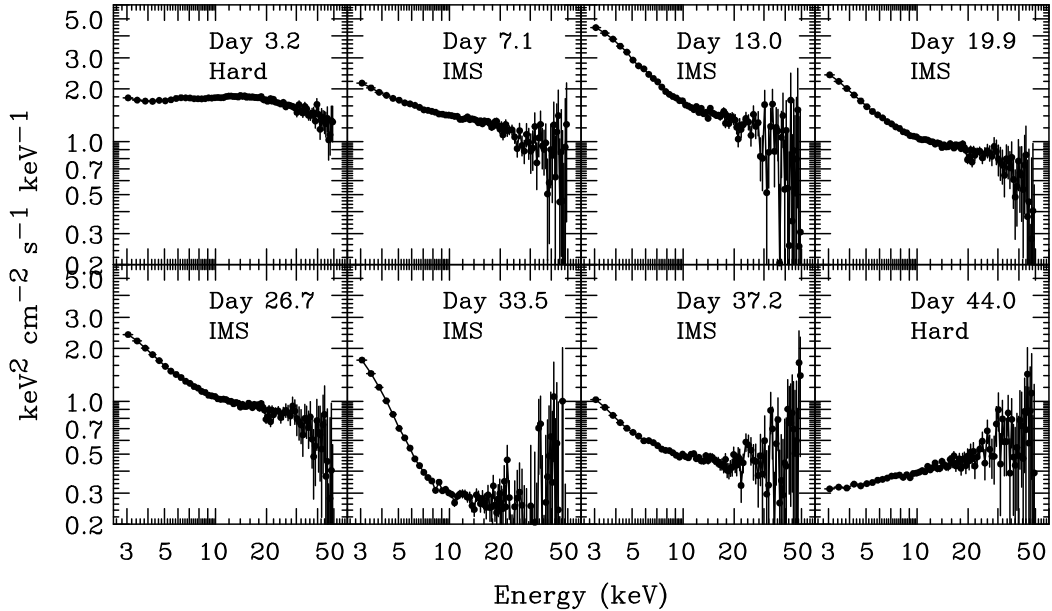}
  \end{center}
  \caption{RXTE/PCA $\nu$F$\nu$ spectra in the 3--50 keV range for typical observations.  The shape of the energy spectra dramatically changed during the outburst. Day indicates days since MJD 55464 (=September 25, 2010).} 
\label{fig2}
\end{figure*}

We first fit the PCA spectra with a disk blackbody (or MCD model; \cite{mcd1}) 
 plus a power-law modified with the Galactic absorption (i.e. {\tt wabs*(diskbb+powerlaw)})
 which are typical components in the X-ray spectra of BHCs.  
 However, a reflection-like structure was found above 7 keV. In addition, a high energy cutoff 
 was required at 30--40 keV in some observations during the initial phase of the outburst. Hence we used the model 
{\tt wabs*(diskbb+smedge*cutoffp)} (hereafter model A) throughout all the data.
The hydrogen column density ($N_{\rm H}$) in {\tt wabs} was fixed at the Galactic value of
 1.7$\times$10$^{21}$ cm$^{-2}$ \citep{nh} following \citet{rxte_timing} 
 since the PCA can not determine 
 $N_{\rm H}$ due to the lack of low energy sensitivity. Using a slightly higher 
 value of 3$\times$10$^{21}$ cm$^{-2}$ (see Section \ref{xrtpcafit}), we found that the fitting results did not change significantly. 
The energy and the width of the smeared edge 
 \citep{gs1124} were fixed at 7.11 keV and 10 keV due to the neutral iron-K edge.
 The averaged ${\chi}^2$ is 56.5 for 73 degrees 
 of freedom (d.o.f.) and the maximum $\chi^2$ is 86.5, suggesting that our 
spectral fits are reasonable. We also tried two other spectral models, 
 {\tt wabs*(diskbb+cutoffp+gaussian)} and {\tt wabs*(diskbb+pexrav)}, 
 using a narrow Gaussian line at 6.4 keV 
 or a reflection component instead of the smeared edge. However, we found that 
 model A gives better fits than these two models 
 (the average $\chi^2$ is 63.3 and 57.9 for 72 d.o.f. for 
 {\tt wabs*(diskbb+cutoffp+gaussian)} and {\tt wabs*(diskbb+pexrav)}, respectively). 
To show the time variation of spectral and timing parameters over the 
 course of the outburst, we selected 8 observations including 2 hard 
 states and 6 IMSs as representatives. Figure \ref{fig2} shows the $\nu F_\nu$ spectra for them. 

\begin{figure}
  \begin{center}
    \FigureFile(80mm,80mm){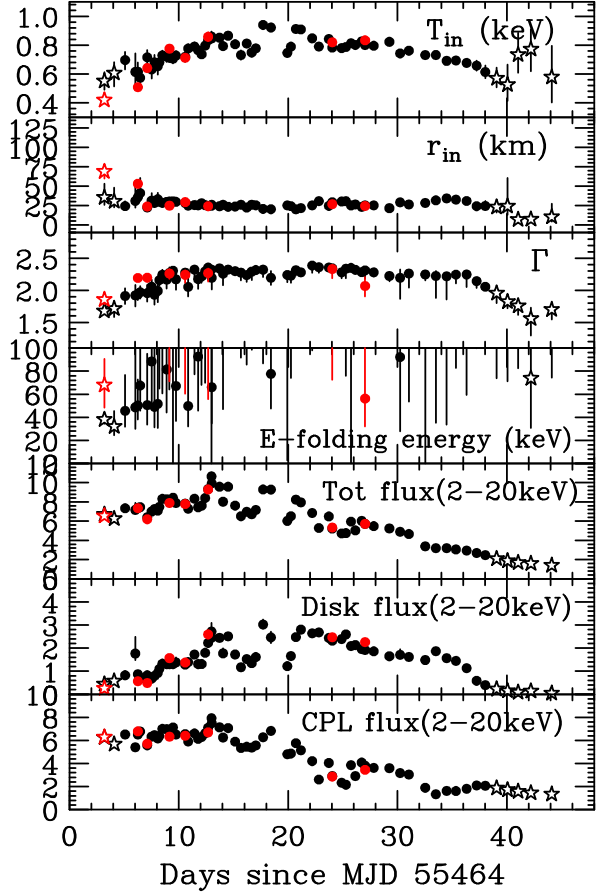}
  \end{center}
  \caption{Time evolution of spectral parameters obtained from model A. The innermost temperature $T_{\rm in}$, innermost radius $r_{\rm in}$, photon index $\Gamma$, e-folding energy, total flux (2--20 keV), disk flux (2--20 keV), and CPL flux  (2--20 keV) are shown from top to bottom. The fluxes are given in units of 10$^{-9}$ erg cm$^{-2}$ s$^{-1}$. The red symbols indicate 8 simultaneous Swift/XRT and RXTE/PCA observations. The hard and intermediate states following definitions in \citet{review1} are shown by open stars and filled circles, respectively. }
\label{fig3}
\end{figure}

Time variation of the spectral parameters (innermost {\tt diskbb} temperature $T_{\rm in}$, 
innermost radius $r_{\rm in}$\footnote{$r_{\rm in}=N_{\rm dbb}^{1/2}(D/10 {\rm kpc})\cos i^{-1/2}$, 
 proportional to the square root of $N_{\rm dbb}$, 
 where $N_{\rm dbb}$ is the normalization of the {\tt diskbb} model, $i$ is the 
 inclination angle (face-on for $i$=0$^{\circ}$), and 
 $D$ is the distance to the source.} (hereafter we assume 
 $D$=10 kpc and $i$=0$^{\circ}$ for simplicity when quoting $r_{\rm in}$), photon index $\Gamma$, 
 and e-folding energy $E_{\rm f}$), the disk flux, 
 the cutoff power-law (CPL) flux, and the total flux in the 2--20 keV range are shown in 
 Figure \ref{fig3}. Note that the observed $T_{\rm in}$ and $r_{\rm in}$ are not physical
  because of the spectral hardening due to the electron scattering in the innermost part 
 of the accretion disk and also the lack of the innermost boundary condition in the MCD model 
 (see Section \ref{discuss2}).
 
 $T_{\rm in}$ increases from $\sim$0.5 to a peak of $\sim$0.8 keV (day 18), 
 and then drops down to 0.5 keV in the decaying phase. 
  During the initial and final phases, where the source is mainly in the hard state, 
  the $r_{\rm in}$ shows some variations with large error bars. 
  Although the fits require the disk component, the disk estimation might have
 large uncertainties due to a low sensitivity below 3 keV of RXTE/PCA (see 
  Section \ref{xrtpcafit}). In fact, we directly compare the $T_{\rm in}$ measured by 
  Swift/XRT published in \citet{swift_dip}, and found that the differences were
   significant during the hard state (0.5--0.6 keV for the PCA versus 
  0.3--0.4 keV for the XRT). 
  $T_{\rm in}$ never exceeds 1 keV, which has been observed in bright BHCs 
 (e.g. GRO J1655--40 
 \citep{gro1655}, H 1743--322 \citep{h1743}, and XTE J1550--564 \citep{xte1550}). 
 In spite of variations of $T_{\rm in}$ and the disk flux, the 
  innermost radius $r_{\rm in}$ remains constant at $\sim$30 km. 

 The $\Gamma$ of the CPL component smoothly varies from 
 1.6 to 2.4 and then back to 1.6. During a few observations in the initial phase, 
 a high energy cutoff at 30--40 keV is required in the fits. This is a bit low compared 
 with typical values of 100--200 keV in the hard state \citep{cutoff}. 
 Some BHCs show a lower cutoff during the initial bright hard state, which is probably 
 due to the rapid increase of the Compton cooling of a high temperature corona by soft disk photons 
 \citep{gx339_lowhard}.  The hypothesis that the coronal temperature is controlled by the disk flux will be discussed later (see Figure \ref{fig8} and \S \ref{discuss1}). 
 The flux of the hard CPL component is independent 
   of the disk-flux variation as generally seen in many BHCs.

\begin{figure}
  \begin{center}
    \FigureFile(80mm,80mm){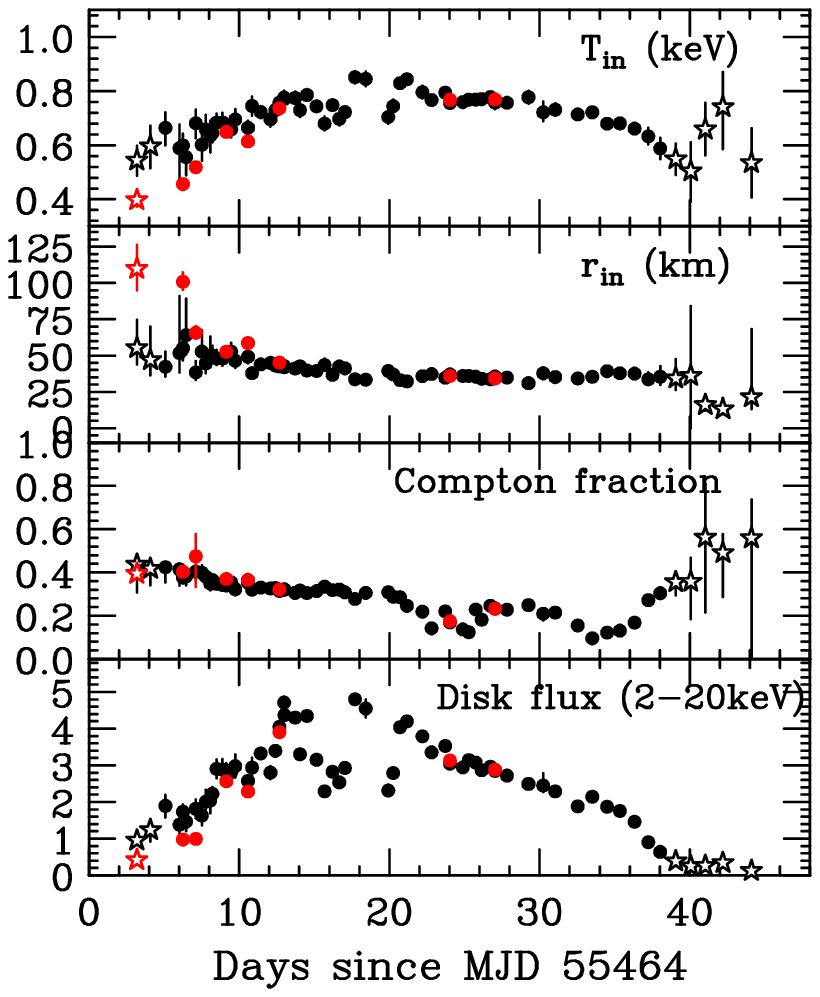}
  \end{center}
  \caption{Time evolution of spectral parameters obtained from model B. The innermost temperature $T_{\rm in}$, innermost radius $r_{\rm in}$, Compton fraction, and disk flux in the 2--20 keV range are shown from top to bottom. The symbols and colors are the same as in Figure \ref{fig3}.}
\label{fig4}
\end{figure}

\citet{simpl} have pointed out that the hard power-law component diverges at low 
 energies in the {\tt wabs*(diskbb+powerlaw)} model. This situation is 
 unphysical because the power-law spectrum results from the Comptonization of disk photons.
 Hence, they developed a self-consistent Comptonization model, {\tt simpl}, in XSPEC. Because the 
 power-law component of MAXI J1659--152 remained relatively strong throughout the outburst, 
 we used this model as a more accurate estimation for the disk parameters (i.e. 
 {\tt wabs*smedge*highecut*simpl$\otimes$diskbb}, 
 where $\otimes$ means a convolution--hereafter model B). We used the 
 up-scattering only option in {\tt simpl}.
 The quality of the fitting with this model was almost the same as in model A, yielding 
 an averaged $\chi^2$ of 56.8 for 73 d.o.f. The time variation of the fitting parameters 
  is shown in Figure \ref{fig4}. Model B fits result in larger innermost 
  radii, $r_{\rm in}$, than those of model A because 
  the {\tt simpl} Comptonization model takes into account the contribution of 
   the disk photons which are scattered into the power-law component as well as 
   direct disk emission.
  The innermost radius is still roughly constant; there is an initial drop from $\sim$50 km
   to $\sim$35 km, where it remains for most of the outburst.

\subsubsection{Simultaneous Broadband Fits between Swift/XRT and RXTE/PCA}\label{xrtpcafit}

  We fit the Swift/XRT and RXTE/PCA spectra simultaneously using models A and B described 
  in Section \ref{pcafit}. 
  To take into account the calibration uncertainties between the two instruments, the
  PCA spectra were normalized to the XRT spectra, multiplying the PCA 
 spctra by a constant factor. We found that this normalization
   factor was reasonably consistent with unity within a 20\% accuracy. 
  The hydrogen column density $N_{\rm H}$ was left free. 

\begin{figure*}
  \begin{center}
    \FigureFile(70mm,70mm){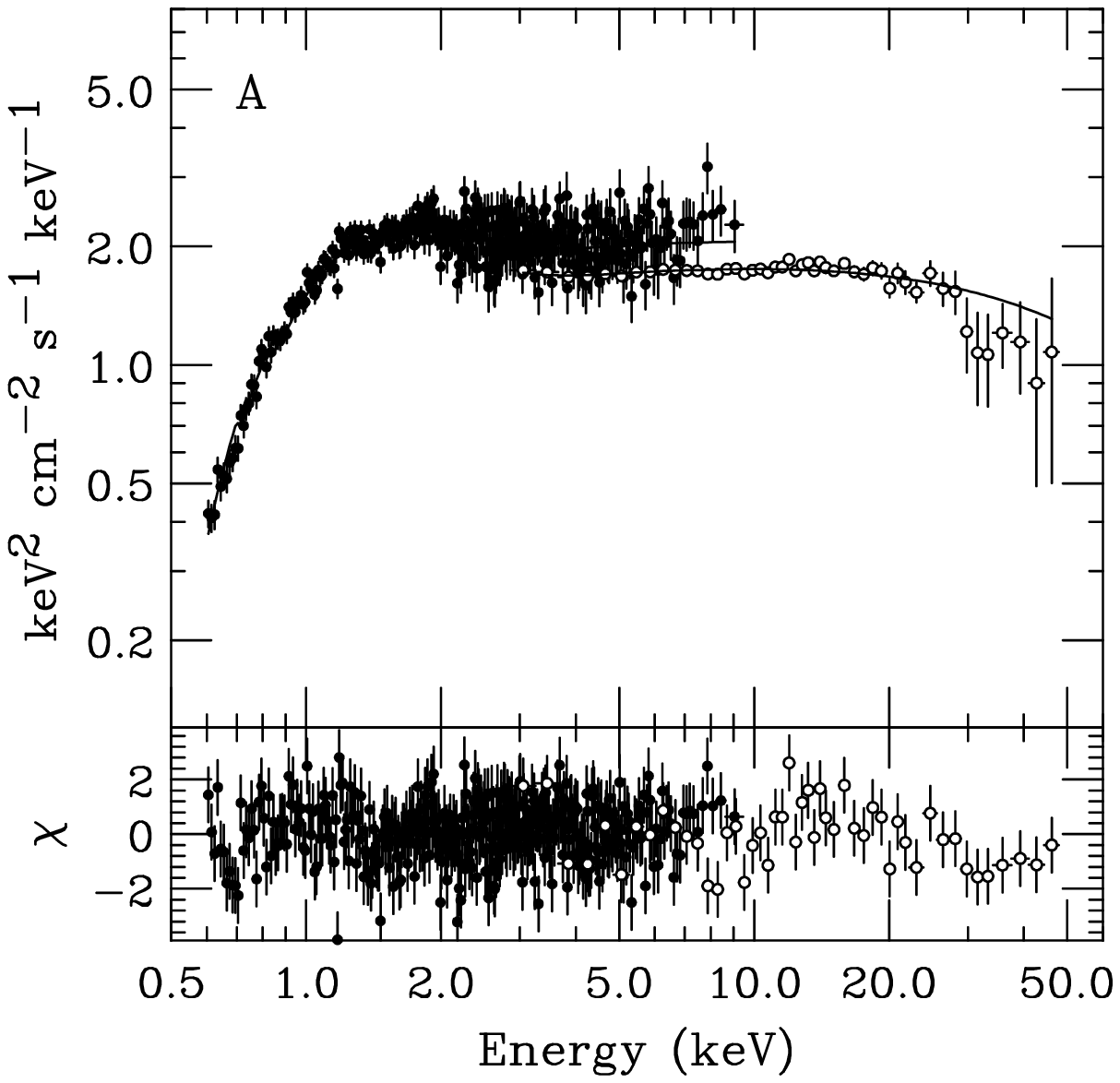}
    \FigureFile(70mm,70mm){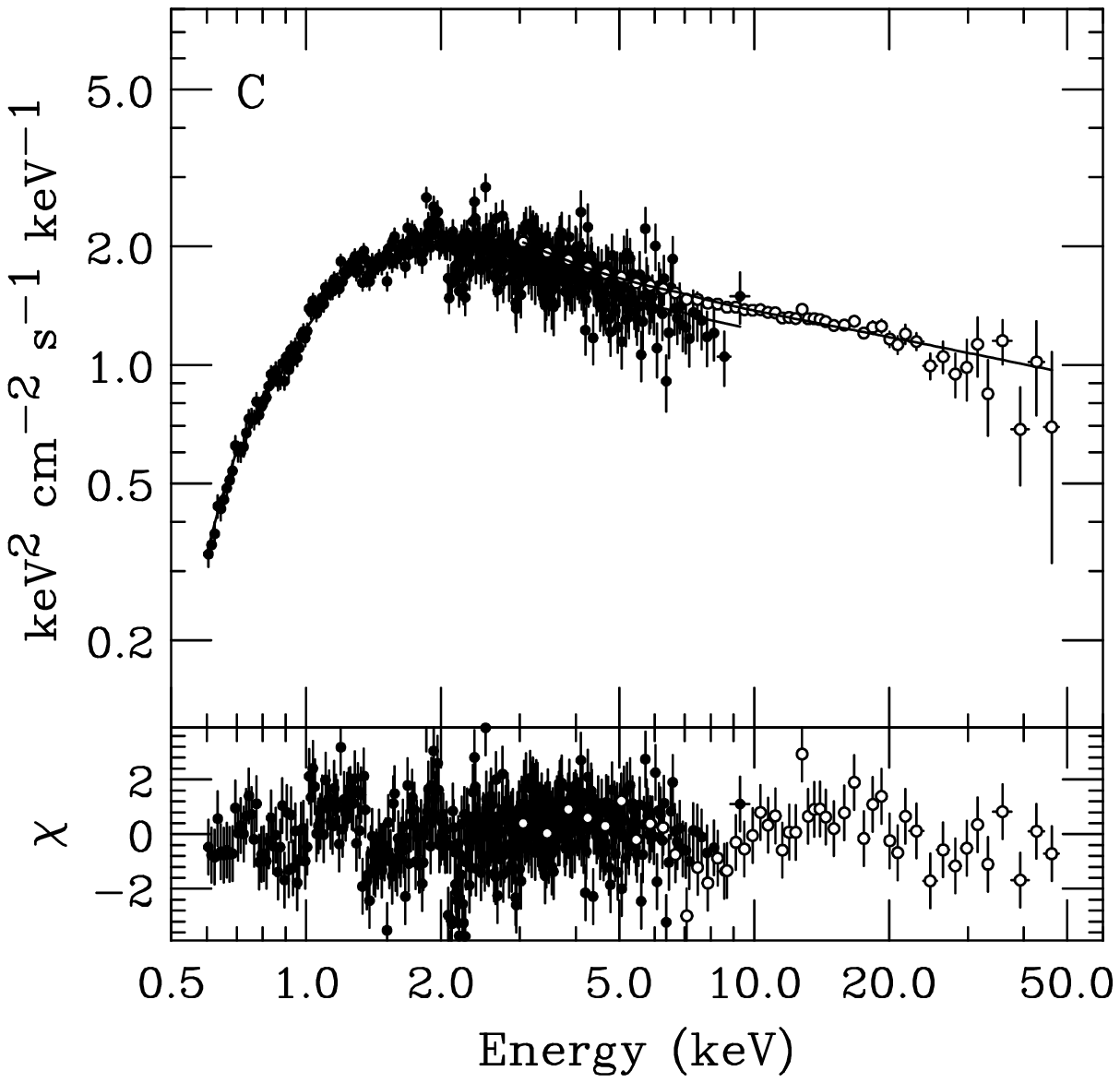}
    \FigureFile(70mm,70mm){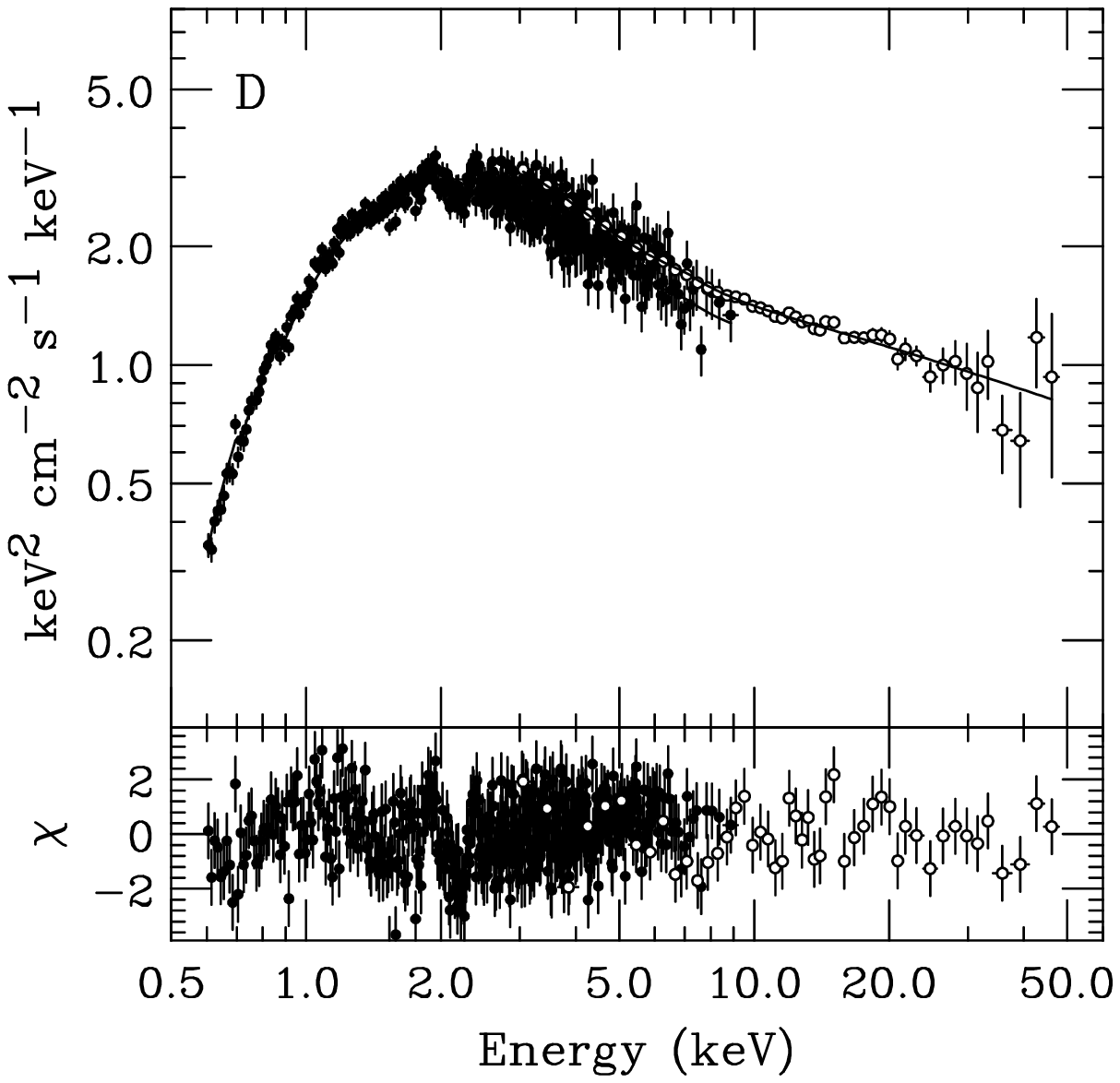}
    \FigureFile(70mm,70mm){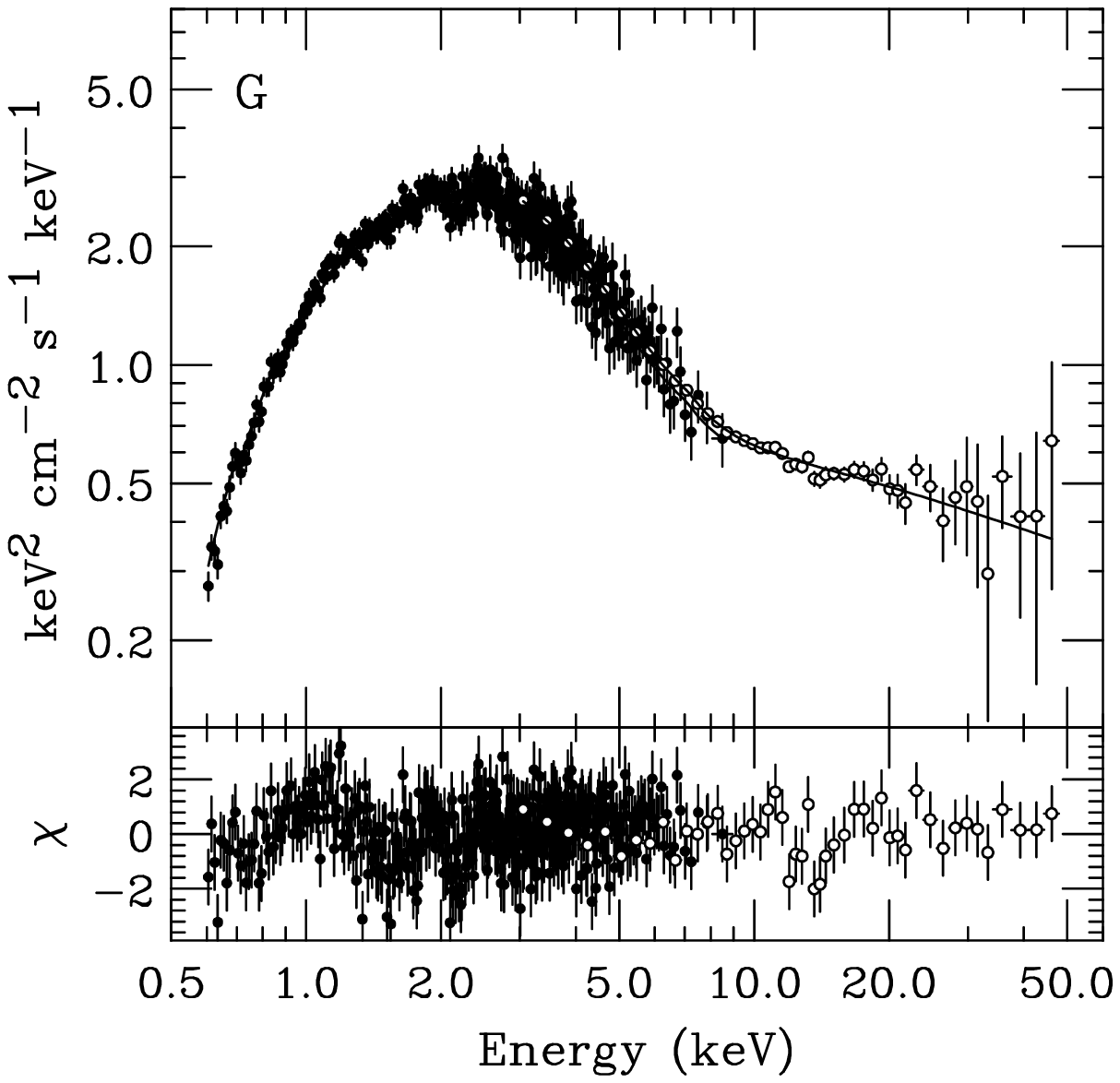}
  \end{center}
  \caption{Spectral fitting results for the Swift/XRT (filled circles) and RXTE/PCA (open circles) data for the period A, C, D, and G (see Table \ref{tab1}). Each upper and lower panel shows the $\nu F_{\nu}$ spectra with the best-fit model B and residuals from model B respectively. }
\label{fig5}
\end{figure*}

 Figure \ref{fig5} shows the best-fit broadband spectra for 4 
 out of 8 simultaneous observation periods.
  The disk blackbody component becomes more evident later in the outburst.
  The fits are reasonable 
 for both models giving reduced chi-squares of 1.0--1.4 for 360 to 500 d.o.f. The best-fit 
 parameters are shown in Table \ref{tab2}. The $N_{\rm H}$ measured in simultaneous fits
  is almost consistent with the Galactic value 0.17$\times$10$^{22}$ cm$^{-2}$ for model B, and model A results in a slightly higher value, by $\sim$0.1$\times$10$^{22}$ cm$^{-2}$, than this value. 
 Their time evolution, shown by red symbols 
 in Figures  \ref{fig3} and \ref{fig4}, displays similar behavior to the results 
 from the RXTE data fits. However, there are significant differences between observations with 
 a lower disk flux than $\sim$1$\times$10$^{-9}$ erg cm$^{-2}$ s$^{-1}$,
 mainly during the hard states; the fits with RXTE data alone tend to give 
 significantly higher $T_{\rm in}$ and smaller $r_{\rm in}$ than those 
 obtained from the Swift+RXTE fits. Thanks to the joint broadband data, we 
 deduce the reduction of $r_{\rm in}$ (by more than a factor of three
  from $\sim$110 km to $\sim$35 km) more precisely than that 
 in the preceding section (from $\sim$50 km to $\sim$35 km). 
 Hereafter, we do not discuss the disk parameters, $T_{\rm in}$ 
 and $r_{\rm in}$, obtained from RXTE observations alone if the observations had lower disk fluxes 
 than 1$\times$10$^{-9}$ erg cm$^{-2}$ s$^{-1}$ in the 2--20 keV range or were classified in the hard state.

\begin{table*}[htbp]
\begin{center}
\caption{List of simultaneous RXTE/PCA and Swift/XRT observations.}\label{tab1}
\begin{tabular}{llcccccc}\\\hline\hline
Period & Instrument & ObsID    & Start Time (UT)     & End Time (UT) & State\footnotemark[$*$] & QPO\footnotemark[$\dagger$]  \\\hline 
A (Day 3.2) & RXTE/PCA  & 95358-01-02-00 & 2010-09-28 00:54:56 & 2010-09-28 08:18:56     & Hard & C\\ 
  & Swift/XRT & 00434928005    & 2010-09-28 07:07:02 & 2010-09-28 07:24:58 &     & \\  
  & Time intervals   &         & 2010-09-28 07:14:41 & 2010-09-28 07:24:55 &     & \\
B (Day 6.3) & RXTE/PCA  & 95358-01-03-00 & 2010-10-01 05:46:40 & 2010-10-01 07:05:04     & IMS & C\\ 
  & Swift/XRT & 00434928009    & 2010-10-01 05:46:02 & 2010-10-01 06:08:59  &  & \\
  & Time intervals  &                  & 2010-10-01 05:49:53 &  2010-10-01 06:08:54 &     & \\
C (Day 7.1) & RXTE/PCA  & 95358-01-03-01 & 2010-10-02 02:32:32 & 2010-10-02 03:37:36     & IMS & C\\ 
  & Swift/XRT & 00434928010    & 2010-10-02 02:38:02 & 2010-10-02 03:00:59 &  & \\
  & Time intervals  &                & 2010-10-02 02:40:43 & 2010-10-02 03:00:54 &  & \\
D (Day 9.1) & RXTE/PCA  & 95108-01-08-00 & 2010-10-04 02:43:44 & 2010-10-04 04:21:52     & IMS & C\\
  & Swift/XRT & 00434928012    & 2010-10-04 02:41:02 & 2010-10-04 03:10:58 &  & \\  
  & Time intervals  &                & 2010-10-04 02:50:41 & 2010-10-04 03:10:32 &     & \\
E (Day 10.6) & RXTE/PCA  & 95108-01-11-00 & 2010-10-05 13:46:24 & 2010-10-05 14:27:44     & IMS & C\\ 
  & Swift/XRT & 00434928013    & 2010-10-05 14:03:02 & 2010-10-05 14:33:01 &  & \\
  & Time intervals  &          & 2010-10-05 14:04:52  & 2010-10-05 14:14:41  &     & \\
F (Day 12.7) & RXTE/PCA  & 95108-01-17-00 & 2010-10-07 16:01:36 & 2010-10-07 16:39:44     & IMS & C\\ 
  & Swift/XRT & 00434928016    & 2010-10-07 15:39:02 & 2010-10-07 16:08:58 &  & \\
  & Time intervals  &                & 2010-10-07 16:01:37  & 2010-10-07 16:08:52 &     & \\
G (Day 24.0) & RXTE/PCA  & 95118-01-05-00 & 2010-10-19 00:25:36 & 2010-10-19 01:17:52     & IMS & no \\ 
  & Swift/XRT & 00031843001    & 2010-10-19 00:51:02 & 2010-10-19 01:15:58 &  & \\   
  & Time intervals  &                & 2010-10-19 01:06:24 & 2010-10-19 01:15:30 &     & \\
H (Day 27.0) & RXTE/PCA  & 95118-01-08-00 & 2010-10-22 00:38:24 & 2010-10-22 01:00:00     & IMS & no \\ 
  & Swift/XRT & 00031843007    & 2010-10-22 00:58:02 & 2010-10-22 01:22:01 &  & \\
  & Time intervals &           & \multicolumn{2}{c}{all intervals for both instruments.} & & \\\hline
\multicolumn{7}{@{}l@{}}{\hbox to 0pt{\parbox{180mm}{\footnotesize 
\hspace{1mm}
\par\noindent
\footnotemark[$*$] Follows definitions in \citet{review1}.  
\par\noindent
\footnotemark[$\dagger$] Follows definitions in \citet{qpo_def2}.
}\hss}}
\end{tabular}
\end{center}
\end{table*}

\begin{table*}
\begin{center}
{\footnotesize
\caption{Best-fit parameters in simultaneous Swift/XRT and RXTE/PCA observations}\label{tab2}
\begin{tabular}{cccccccccc}\\\hline\hline
Model & Para. & \multicolumn{2}{c}{Period A (Day 3.2)} & \multicolumn{2}{c}{Period B (Day 6.3)} & \multicolumn{2}{c}{Period C (Day 7.1)} & \multicolumn{2}{c}{Period D (Day 9.1)}  \\
      &       & model A & model B & model A & model B & model A & model B & model A & model B \\\hline
wabs   & N$_{\rm H}$\footnotemark[$*$] 	& 0.27$^{+0.01}_{-0.02}$ & 0.23$^{+0.02}_{-0.02}$	& 0.31$^{+0.01}_{-0.01}$ & 0.23$^{+0.01}_{-0.01}$	& 0.29$^{+0.01}_{-0.01}$ & 0.18$^{+0.01}_{-0.01}$	& 0.34$^{+0.02}_{-0.02}$ & 0.21$^{+0.01}_{-0.01}$	\\
diskbb & $T_{\rm in}$(keV)	& 0.42$^{+0.03}_{-0.02}$ & 0.40$^{+0.03}_{-0.03}$	& 0.51$^{+0.01}_{-0.01}$ & 0.46$^{+0.01}_{-0.01}$	& 0.64$^{+0.02}_{-0.02}$ & 0.52$^{+0.02}_{-0.02}$	& 0.78$^{+0.02}_{-0.02}$ & 0.65$^{+0.01}_{-0.01}$	\\
       & $r_{\rm in}$(km)	& 69.0$^{+8.9}_{-9.2}$ & 109.7$^{+16.6}_{-14.9}$	& 53.1$^{+3.9}_{-3.6}$ & 100.8$^{+6.3}_{-5.7}$	& 23.8$^{+2.6}_{-2.4}$ & 65.7$^{+4.6}_{-4.0}$	& 24.7$^{+2.3}_{-2.3}$ & 52.6$^{+2.0}_{-1.9}$	\\
cutoffp/ & $\Gamma$	& 1.86$^{+0.05}_{-0.07}$ & 1.87$^{+0.04}_{-0.06}$	& 2.19$^{+0.01}_{-0.05}$ & 2.22$^{+0.01}_{-0.02}$	& 2.20$^{+0.02}_{-0.05}$ & 2.24$^{+0.02}_{-0.02}$	& 2.26$^{+0.07}_{-0.08}$ & 2.39$^{+0.02}_{-0.02}$	\\
simpl        & $E_{\rm f}$(keV)\footnotemark[$\dagger$] 	& 68.0$^{+22.7}_{-19.7}$ & 73.4$^{+32.5}_{-20.6}$	& $>$176.3 & ---	& $>$179.3 & ---	& $>$61.2 & ---	\\
             & $A$\footnotemark[$\ddagger$]  	& 1.46$^{+0.07}_{-0.13}$ & --- 	& 2.64$^{+0.08}_{-0.17}$ & --- 	& 2.24$^{+0.08}_{-0.15}$ & --- 	& 2.88$^{+0.29}_{-0.30}$ & --- 	\\
             & $f$\footnotemark[$\S$]  	& --- & 0.39$^{+0.02}_{-0.02}$	& --- & 0.40$^{+0.01}_{-0.02}$	& --- & 0.47$^{+0.10}_{-0.14}$	& --- & 0.37$^{+0.01}_{-0.01}$	\\
smedge & $\tau$\footnotemark[$\|$] 	& $<$0.27 & $<$0.21	& $<$0.20 & $<$0.14	& $<$0.27 & $<$0.28	& 0.34$^{+0.22}_{-0.22}$ & 0.47$^{+0.15}_{-0.16}$	\\
constant & factor	& 1.17$^{+0.02}_{-0.02}$ & 1.17$^{+0.02}_{-0.02}$	& 0.94$^{+0.01}_{-0.01}$ & 0.94$^{+0.01}_{-0.01}$	& 0.90$^{+0.01}_{-0.01}$ & 0.89$^{+0.01}_{-0.01}$	& 0.87$^{+0.01}_{-0.01}$ & 0.86$^{+0.01}_{-0.01}$	\\
$\chi^2$/d.o.f. & 	& 500.1/425 & 502.2/425	& 669.4/482 & 687.8/482	& 645.6/474 & 682.7/474	& 645.6/500 & 691.2/500	\\\hline
Model & Para. & \multicolumn{2}{c}{Period E (Day 10.6)} & \multicolumn{2}{c}{Period F (Day 12.7)} & \multicolumn{2}{c}{Period G (Day 24.0)} & \multicolumn{2}{c}{Period H (Day 27.0)} \\
      &       & model A & model B & model A & model B & model A & model B & model A & model B \\\hline
wabs   & N$_{\rm H}$\footnotemark[$*$] 	& 0.33$^{+0.01}_{-0.03}$ & 0.21$^{+0.01}_{-0.01}$	& 0.33$^{+0.02}_{-0.03}$ & 0.19$^{+0.01}_{-0.01}$	& 0.30$^{+0.01}_{-0.02}$ & 0.20$^{+0.01}_{-0.01}$	& 0.32$^{+0.02}_{-0.02}$ & 0.24$^{+0.01}_{-0.01}$	\\
diskbb & $T_{\rm in}$(keV)	& 0.71$^{+0.02}_{-0.02}$ & 0.61$^{+0.01}_{-0.01}$	& 0.86$^{+0.03}_{-0.03}$ & 0.74$^{+0.02}_{-0.02}$	& 0.82$^{+0.01}_{-0.01}$ & 0.77$^{+0.01}_{-0.01}$	& 0.83$^{+0.02}_{-0.01}$ & 0.77$^{+0.01}_{-0.01}$	\\
       & $r_{\rm in}$(km)	& 29.4$^{+3.5}_{-3.4}$ & 58.7$^{+3.2}_{-2.9}$	& 24.2$^{+2.7}_{-2.3}$ & 45.1$^{+2.5}_{-2.4}$	& 26.7$^{+1.3}_{-0.8}$ & 35.9$^{+1.1}_{-1.1}$	& 24.4$^{+1.5}_{-1.5}$ & 34.3$^{+1.1}_{-1.1}$	\\
cutoffp/ & $\Gamma$	& 2.24$^{+0.07}_{-0.12}$ & 2.35$^{+0.03}_{-0.03}$	& 2.27$^{+0.07}_{-0.13}$ & 2.41$^{+0.04}_{-0.04}$	& 2.33$^{+0.03}_{-0.14}$ & 2.40$^{+0.04}_{-0.04}$	& 2.07$^{+0.14}_{-0.16}$ & 2.35$^{+0.04}_{-0.04}$	\\
simpl    & $E_{\rm f}$(keV)\footnotemark[$\dagger$] 	& $>$60.3 & ---	& $>$55.7 & ---	& $>$72.6 & ---	& 56.2$^{+103.4}_{-24.2}$ & ---	\\
         & $A$\footnotemark[$\ddagger$]  	& 2.81$^{+0.21}_{-0.44}$ & --- 	& 3.08$^{+0.43}_{-0.49}$ & --- 	& 1.42$^{+0.11}_{-0.25}$ & --- 	& 1.21$^{+0.24}_{-0.24}$ & --- 	\\
         & $f$\footnotemark[$\S$]  	& --- & 0.37$^{+0.02}_{-0.02}$	& --- & 0.32$^{+0.01}_{-0.01}$	& --- & 0.17$^{+0.01}_{-0.01}$	& --- & 0.23$^{+0.01}_{-0.01}$	\\
smedge & $\tau$\footnotemark[$\|$] 	& 0.45$^{+0.30}_{-0.29}$ & 0.48$^{+0.21}_{-0.21}$	& 0.60$^{+0.36}_{-0.37}$ & 0.89$^{+0.24}_{-0.24}$	& 0.47$^{+0.35}_{-0.28}$ & 0.81$^{+0.25}_{-0.25}$	& 0.81$^{+0.38}_{-0.38}$ & 0.84$^{+0.26}_{-0.26}$	\\
constant & factor	& 0.97$^{+0.02}_{-0.02}$ & 0.96$^{+0.02}_{-0.02}$	& 1.09$^{+0.02}_{-0.02}$ & 1.09$^{+0.02}_{-0.02}$	& 0.93$^{+0.01}_{-0.01}$ & 0.93$^{+0.01}_{-0.01}$	& 0.87$^{+0.01}_{-0.01}$ & 0.87$^{+0.01}_{-0.01}$	\\
$\chi^2$/d.o.f. & 	& 457.6/409 & 482.9/409  & 366.6/363 & 379.6/363	& 549.7/462 & 562.1/462	& 646.5/495 & 642.0/495	\\\hline
\multicolumn{10}{@{}l@{}}{\hbox to 0pt{\parbox{180mm}{\footnotesize 
\hspace{1mm}
\par\noindent
\footnotemark[$*$] In unit of 10$^{22}$ cm$^{-2}$.  
\par\noindent
\footnotemark[$\dagger$] An e-folding energy in {\tt cutoffp} model or {\tt highecut}.  
\par\noindent
\footnotemark[$\ddagger$] A normalization of {\tt cutoffp} in unit of photons cm$^{-2}$ s$^{-1}$ at 1 keV.
\par\noindent
\footnotemark[$\S$] A Compton fraction in {\tt simpl}.
\par\noindent
\footnotemark[$\|$] An optical depth for the neutral iron K-edge at 7.11 keV.
}\hss}}
\end{tabular}
}
\end{center}
\end{table*}

\subsection{PCA Timing results}

\begin{figure*}
  \begin{center}
    \FigureFile(140mm,140mm){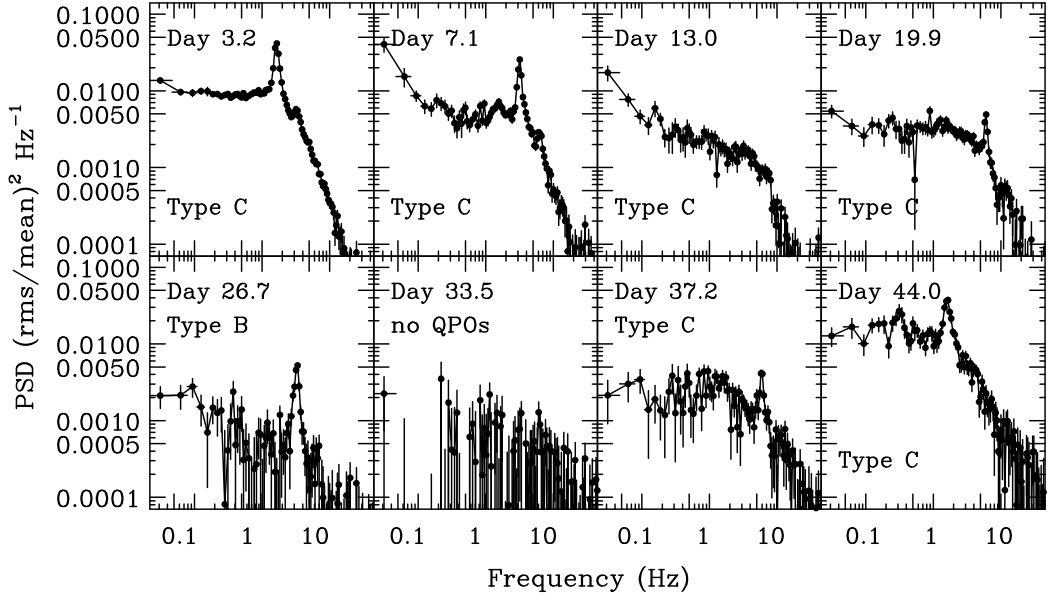}
 \end{center}
 \caption{The power density spectra (PDS) for the 8 representative observations shown in Figure \ref{fig2}. These are produced using the full PCA energy range (2--100 keV). Day means days since MJD 55464 (= September 25, 2010). }\label{fig6}
\end{figure*}

\begin{figure}
  \begin{center}
    \FigureFile(80mm,80mm){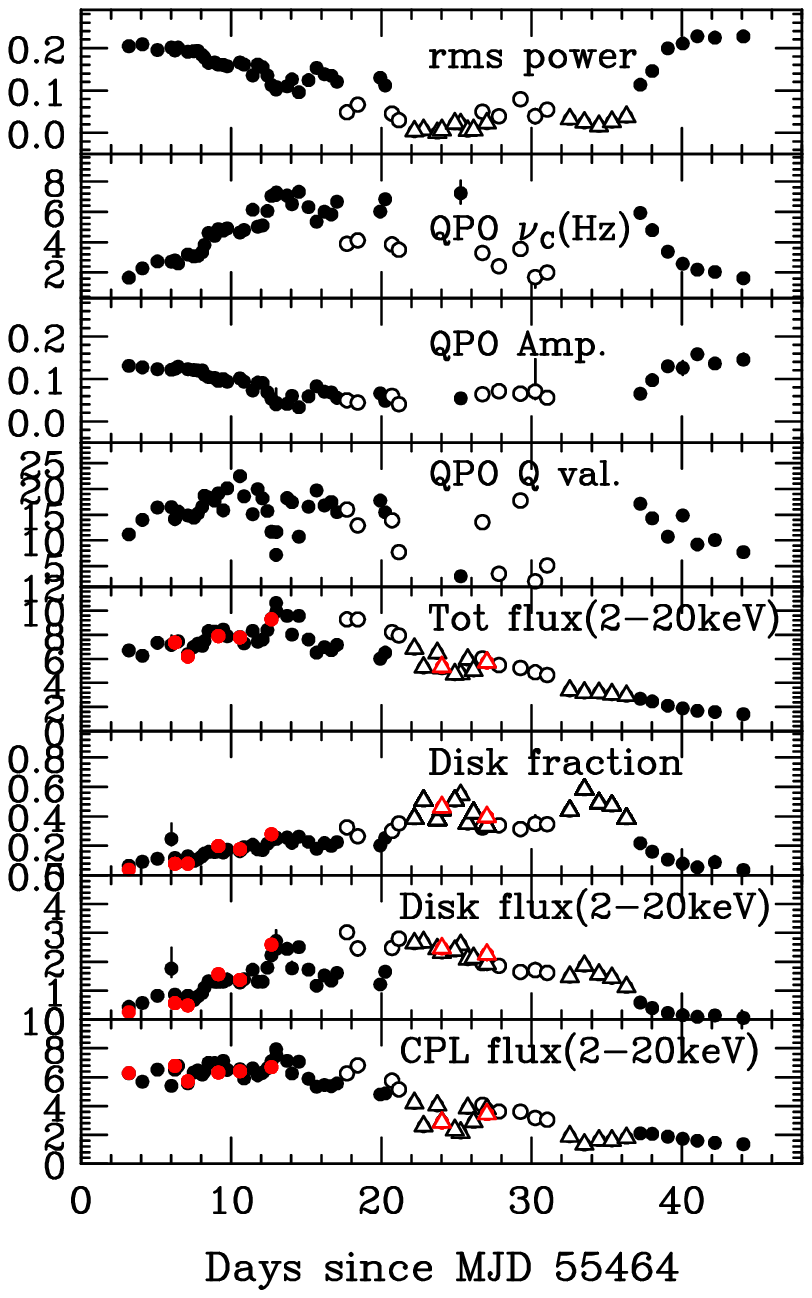}
  \end{center}
  \caption{Time evolution of timing parameters. The rms variability (integrated over 0.1--10 Hz), QPO central frequency $\nu_{\rm c}$, QPO amplitude $A$, $Q$ value, total flux (2--20 keV), disk fraction, disk flux (2--20 keV), and CPL flux (2--20 keV) are shown from top to bottom. The open circles, filled circles, and triangles  denote observations which showed type-B QPO, type-C QPOs, and no QPOs, respectively.}
\label{fig7}
\end{figure}

Figure \ref{fig6} shows the power spectra for 8 representative observations\footnote{Please note that these 8 observations are not identical to the 8 simultaneous XRT and PCA observations.} as displayed by arrows in Figure \ref{fig1}. 
To derive the QPO characteristics, the power spectra were modeled with a single
 Lorentzian plus a power-law in a narrow frequency range around the QPO, {\it i.e.}, 

\begin{equation}
 P(\nu) = \frac{A^2}{\Delta \nu\pi} \frac{1}{1+(\frac{{\nu}-{\nu}_{\rm c}}{\Delta \nu})^2} + B{\nu}^{-\alpha}
\end{equation}

where ${\nu}_{\rm c}$ is the QPO central frequency, $\Delta \nu$ is the FWHM width, and 
 $A$ is the QPO amplitude integrated over the full frequency range.  
 The coherence parameter $Q$ is defined by $\nu_{\rm c}$/$\Delta \nu$. We focus 
 on the low frequency QPOs (LFQPOs) in these fits because no significant high 
 frequency QPOs were detected.  We follow results from \citet{rxte_spectiming} for the
 classification of QPOs (Type A, B and C; \cite{qpo_def2}) 
 between 1 and 8 Hz in 51 of 65 observations; type-B and C were observed in 
 9 and 42 observations, respectively. 
 The results of  central frequency, widths, and the rms variability
  in the QPO parameters are consistent with \citet{rxte_spectiming}.
 Figure \ref{fig7} shows the LFQPO evolution with time. At the beginning of the outburst, 
 the frequency of the type-C QPOs gradually 
 increased from 1.6 to 7.3 Hz and then disappeared when the disk 
 fraction was larger than 0.26. Later into the outburst, 
 the type-C QPOs appeared again at 5.9 Hz 
 and decreased down to 1.6 Hz. Type-B QPOs were 
 seen at 1.7--4.1 Hz in the limited time period when the disk fraction was a bit higher than 
 in the case of type-C QPOs. The red noise component under the Type-B QPO is roughly explained by 
 $\nu^{-\alpha}$, where $\alpha$ ranges between 0.7--1.1. This is consistent with 
 ${\nu}^{-1}$, as typically seen in the thermal state (e.g. \cite{review2}).

\subsection{Correlation among spectral and timing parameters}

\begin{figure*}
  \begin{center}
    \FigureFile(170mm,170mm){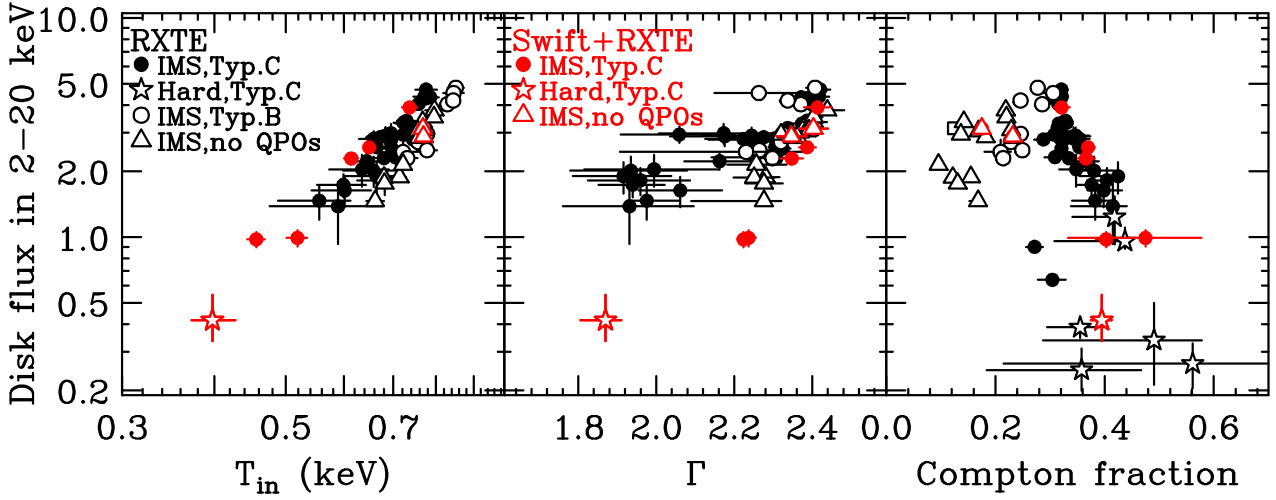}
  \end{center}
  \caption{Correlation between the disk flux in 2--20 keV range and other spectral parameters (innermost temperature $T_{\rm in}$, photon index $\Gamma$, and the Compton fraction.). The results are based on spectral fitting with the model B. The same symbols are used for Figure \ref{fig7}.}
\label{fig8}
\end{figure*}

 To clarify the origin of LFQPOs, we investigated the correlation (or lack thereof) 
 between obtained spectral and timing parameters. Figure \ref{fig8} shows the correlation of 
 the disk flux in the 2--20 keV range with innermost temperature $T_{\rm in}$, photon index 
 $\Gamma$, and Compton fraction. We can see a clear correlation between the disk flux and both
 $T_{\rm in}$ and $\Gamma$, while the Compton fraction is 
 independent of the disk flux. The middle panel (a steeper photon index for a 
   more luminous disk) suggests that the coronal temperature is controlled by the disk flux 
  (see Section \ref{discuss1}). 
 
\begin{figure*}
  \begin{center}
    \FigureFile(170mm,170mm){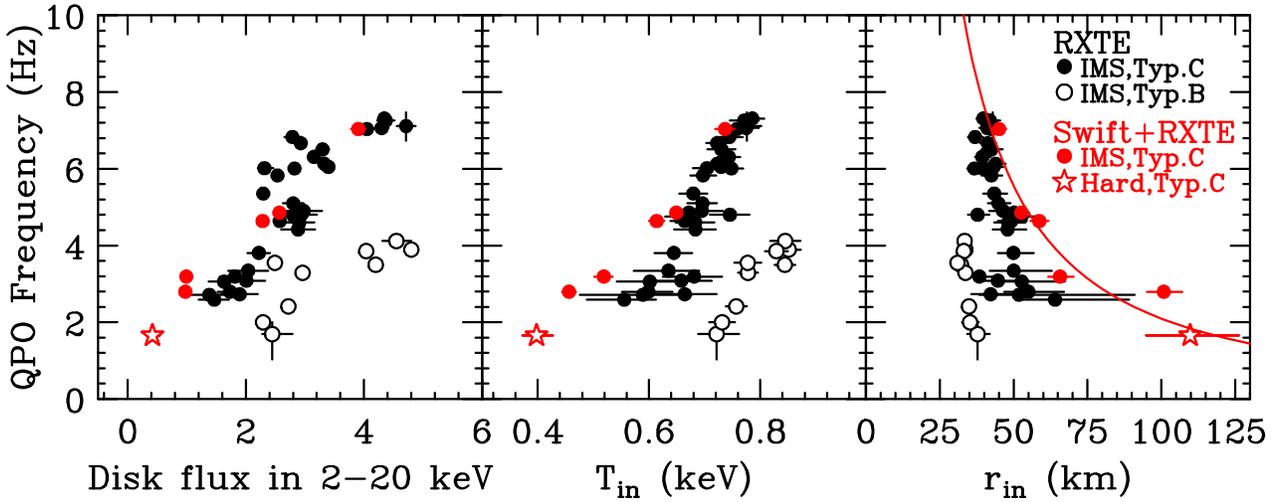}
  \end{center}
  \caption{Correlation between QPO frequency and spectral parameters (disk flux, innermost temperature, and innermost radius) derived from model B. The red line shows the best-fit model ($r_{\rm in}\propto \nu_{\rm c}^{-0.70}$) for simultaneous Swift/XRT and RXTE/PCA data. The type-B QPOs, and type-C QPOs during the IMS, and type-C QPOs during the hard state indicates the open circles, filled circles, and open stars, respectively. The red symbol indicates the joint Swift/RXTE observations. }
\label{fig9}
\end{figure*}

\begin{figure*}
  \begin{center}
    \FigureFile(170mm,170mm){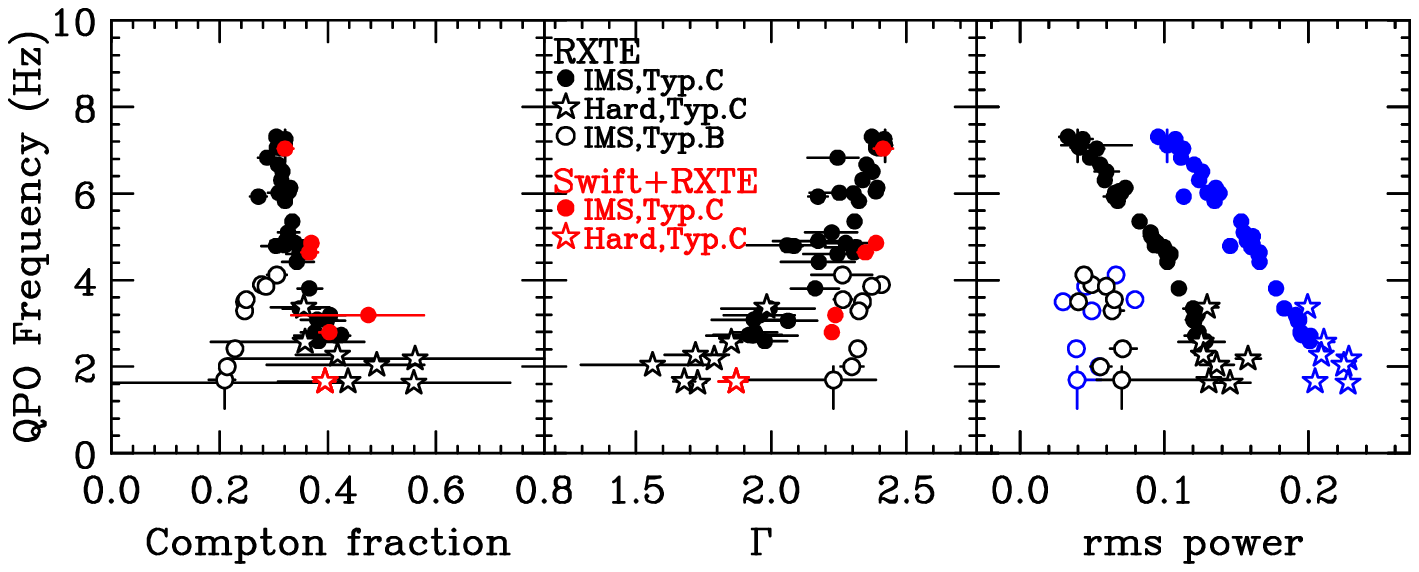}
  \end{center}
  \caption{Correlation between QPO frequency and spectral parameters (II) (Compton fraction and the photon index $\Gamma$, the QPO amplitude $A$ (black), and the rms variability integrated over 0.1--10 Hz (blue). The symbols are the same as in Figure \ref{fig8}. }
\label{fig10}
\end{figure*}

 Figures \ref{fig9} and \ref{fig10} show the QPO frequency $\nu_{\rm c}$ as a function of 
 the disk flux (2--20 keV), $T_{\rm in}$, $r_{\rm in}$, Compton fraction, 
  $\Gamma$, QPO amplitude $A$, and rms variability (integrated over 0.1--10 Hz).
 There is a positive correlation between the type-C QPO $\nu_{\rm c}$ and both the disk flux 
 and $T_{\rm in}$. We also confirmed the $\nu_{\rm c}$-$\Gamma$ correlation already suggested by 
 \citet{rxte_spectiming2}. The correlation is negative with Compton fraction, 
 QPO amplititude $A$, and fractional rms variability (including the noise component). The slopes of 
 the $\nu_{\rm c}$-$A$ and $\nu_{\rm c}$-rms correlations during the type-C QPO detections 
  are almost the same, i.e. $A$=$-$0.017$\pm0.001\nu_{\rm c}$(Hz)+0.171$\pm$0.005 and 
  rms=$-$0.019$\pm0.001\nu_{\rm c}$(Hz)+0.248$\pm$0.002.
 The $r_{\rm in}$ obtained from model B shows an anti-correlation with $\nu_{\rm c}$. 
  This can be explained by $r_{\rm in}({\rm km}) = 167^{+33}_{-29} \nu_{\rm c}({\rm Hz})^{-0.70 
 \pm 0.12}$ for combined XRT and PCA data.
  All of the correlations obtained by PCA data alone can be confirmed by simultaneous Swift/XRT and
  RXTE/PCA observations.
 We also found that the correlation of type-C $\nu_{\rm c}$ does not depend on which 
 state the source was in. This implies that although we have classified the 
 source into two separate states (IMS and hard state), the QPO may have the same physical 
  origin in the different states.  
 
 The type-B QPOs show different behavior than the type-C QPOs in the correlation plots.
 There are no correlations of $\nu_{\rm c}$ with $r_{\rm in}$,  $\Gamma$, and 
  rms variability. During the type-B QPOs, the innermost radius 
   $r_{\rm in}$ is consistent with a constant value, independent of $\nu_{\rm c}$,
    of 33.4$\pm$1.2 km (the constant fit gives 
  $\chi^2$/dof = 3.95/8). This radius could possibly correspond to the ISCO.
 
\section{Discussions}\label{discuss}

\subsection{Short Summary and Comparison with Previous Results and Other Black Hole Binaries}\label{discuss1}

 We have presented detailed spectral and timing results from the 2010 outburst of a new BHC, 
 MAXI J1659--152. We used the {\tt simpl} Comptonization model for spectral modeling, 
  and mainly investigated a correlation between spectral parameters and the QPO parameters.

 According to the classification criteria given by \citet{review1}, the source made a state 
 transition into the intermediate state from the hard state 
 in the early phase and went back to the hard state during the final phase of the outburst.
  It stayed in the intermediate state during almost of all the observations and never 
 entered into the SPL state or thermal state.
 This fact is already mentioned by \citet{rxte_timing}.
 The RXTE observations revealed that the time evolution of spectral and timing parameters 
 in the 2010 outburst of MAXI J1659--152 are similar to other BHCs (e.g. \cite{review2}). 
 The PCA spectra in the 3--50 keV range and XRT+PCA broadband spectra in the 0.6--50 keV
 range were successfully fit by both the MCD \citep{mcd1} plus a power-law with an exponential 
 cutoff and the MCD convolved with {\tt simpl} Comptonization \citep{simpl}. 
 The maximum $T_{\rm in}$ in the accretion disk was $\sim$0.9 keV, which is a bit 
 lower than the typical $\sim$1.2 keV seen in bright BHCs such as GRO J1655--40 \citep{gro1655} 
 and XTE J1550--564 \citep{xte1550}. 
 Since $T_{\rm in}$ is proportional to $\beta^{-1/2}\eta^{1/4} M^{-1/4}$ \citep{hightin}, 
 the Eddington ratio ($\eta$=$L/L_{\rm E}$) is relatively low if the BH mass 
 $M$ and spin (related to $\beta$=$R_{\rm in}$/3$R_{\rm S}$; see Section \ref{discuss2}) 
 are similar to these bright sources. During the intermediate states, the innermost 
 radius remained almost constant at $\sim$30 km (35 km in {\tt simpl} fits) in spite of 
 large variations in the disk flux. The flux variation of the hard CPL component was 
 independent of the disk flux. The photon index in the hard component ranged from 1.6 
 to 2.4, in correlation with the disk flux, and did not reach typical 
 SPL values of 2.5--3.0 \citep{review2}.  
 
 In the initial phase, a high energy cutoff in the power-law component 
 was visible at 30--40 keV. The cutoff energy then evolved into higher 
 energies than the PCA maximum energy (50 keV).  The observed e-folding energy 
 is much lower than a typical value in the hard state (100--200 keV; \cite{cutoff}), 
 but a low cutoff energy has been seen for several BHCs 
 during the rising phase \citep{gx339_lowhard}, probably due to Compton cooling in the corona 
  by soft photons. This could also indicate a transition of the electron distribution from 
 thermal to non-thermal electrons in the corona, but we cannot investigate this issue 
  due to the limited PCA energy range up to 50 keV.    

  We confirmed results from \citet{rxte_timing} and \citet{rxte_spectiming} 
  that LFQPOs, consisting of type-B and type-C, were observed at 1--8 Hz in 51 of 65 
 observations. The LFQPOs were seen when both disk and corona components were firmly present, and 
 the disk fraction was less than 35 \% (i.e. the hard CPL component was strong). 
 From detailed fits with the {\tt simpl} Comptonization model, we have shown that 
 the type-C QPO central frequency varies from 1.6 Hz to 7.3 Hz in association 
 with the disk flux, the innermost temperature, the innermost radius, the photon index, 
 and the rms variability, similar to that seen in several BHCs (e.g. \cite{index_freq}, \cite{qpofreq_gamma}).  
   
Various types of QPOs including low-frequency and high-frequency QPOs 
 have been studied in many X-ray binaries from cataclysmic variables
 to black holes (\cite{qpo_bh}, \cite{qpo_bh2}, \cite{qpo_cvnsbh}), but their origin is still unknown. 
 Several models such as the Lense-Thirring precession model 
 (\cite{precession}) and the disk oscillation model (\cite{diskosci}) have been proposed. 
 For type-C LFQPOs, we propose a possible scenario based upon the idea of 
 a truncated accretion disk.
 We assume that the high temperature, corona-like, advection dominated 
 accretion flow (ADAF) is well inside the truncated disk \citep{adaf}.
  When the local mass accretion rate from the outer part of the disk increases, the innermost 
 temperature ($T_{\rm in}$) and the disk flux ($\propto T_{\rm in}^4$) also increase 
 (see left panel of Figure \ref{fig8}). 
 As the disk flux increases, the corona undergoes cooling due to the Compton scattering 
 of soft photons from the thin disk, and the Compton $y$ parameter
  (=$\frac{4kT_{\rm e}}{m_{\rm e}c^2}{\rm Max}(\tau, \tau^2$)) gets lower (i.e. the photon index 
 gets steeper; see middle panel of Figure \ref{fig8}). 
 The innermost radius of the disk is determined by a balance between the 
  accretion rate and the gas evaporation rate \citep{evap}, 
  and so the innermost radius approaches the ISCO (3$R_{\rm S}$ (see Section \ref{discuss2})
  corresponding to $r_{\rm in}\sim$35 km) from around 9.4$R_{\rm S}$ ($\sim$110 km) due to a 
 gradual increase of the mass accretion rate.

 \citet{rxte_spectiming2} showed that the type-C QPO rms variability increased with 
 X-ray energy for MAXI J1659--152. Several authors have derived the same results for other BHCs 
 (e.g. \cite{qpo_def2}), naturally implying that the QPO could be of coronal origin.  
 The observed LFQPO frequency (1--8 Hz) is too slow to be explained by the dynamical 
 time scale $t_{\rm dyn}$ (eq.(3.4) in \cite{tim_vis2}:
\begin{equation}
f_{\rm dyn}=\frac{1}{t_{\rm dyn}}=\frac{v_{\phi}}{2\pi r}=366 \left( \frac{M}{6 M_{\odot}} \right)^{-1} \left( \frac{r}{3R_{\rm S}} \right)^{-1.5} \rm Hz
\end{equation}
where $r$ is the distance to the BH and $v_{\phi}$ is the Keplerian velocity),  
 but can be explained by the viscous time scale $t_{\rm vis}$ (\cite{tim_vis1}, {eq.(3.71)} in \cite{tim_vis2}:
\begin{eqnarray}
f_{\rm vis}&= \frac{1}{t_{\rm vis}} =\alpha \left(\frac{h}{r} \right)^2 f_{\rm dyn} \nonumber \\
 &= 11.0 \left(\frac{\alpha}{0.03} \right) \left(\frac{h}{r} \right)^2 \left(\frac{M}{6 M_{\odot}} \right)^{-1} \left(\frac{r}{3R_{\rm S}}\right)^{-1.5} \rm Hz,
\end{eqnarray}
 where $\alpha$ is the viscous parameter and $h$ is the height 
 of the corona). Here, we adopt $\alpha$ to be 0.03 (\cite{alpha1}, \cite{alpha2}, \cite{alpha3}), 
 the scale height of the corona ($h/r$) to be 1, and the BH mass to be 6 solar mass 
 (see Section \ref{discuss2}).  Hence, the corresponding frequency to the radius range of 
  3 to 9.4$R_{\rm S}$, 2.0--11.0 Hz, is in a good agreement 
 with the observed frequency at 1--8 Hz.
 
 If the type-C QPO originates from the neighborhood of the disk truncation radius, 
 we can explain the change of the QPO frequency by a change of the innermost radius. In fact, 
  the observed $r_{\rm in}$-$\nu_{\rm c}$ correlation ($r_{\rm in} \propto 
 \nu_{\rm c}^{-0.70\pm0.12}$) is remarkably consistent with the above relation 
 $\nu_{\rm c} \propto r_{\rm in}^{-1.5}$. A larger innermost radius 
  might result in a larger coronal size, that is, a 
  larger contribution of the hard CPL component, which has more variability than 
  the disk emission in the X-ray spectrum. 
  This could explain the anti-correlation between the QPO frequency and rms variability 
  as seen in right panel of Figure \ref{fig10}.

 In contrast, the frequency of type-B QPOs varied from 1.7 to 4.1 Hz, while the
  innermost radius $r_{\rm in}$ remained constant at $\sim$35 km. The constancy of $r_{\rm in}$
   suggests that the inner edge of the accretion disk had already reached the ISCO during the type-B QPOs.
 Thus, a change of the frequency cannot be explained by a change of the disk innermost radius. 
 An innermost-radius independent mechanism (e.g. magnetic reconnection above the thin disk) 
  is necessary for an explanation of type-B QPOs. 

 The 2010 outburst of MAXI J1659--152 lasted for about two months,
 a much shorter duration than many other black hole systems (typically more than 
 100 days; \cite{review2}). This short duration is consistent with the fact that MAXI J1659--152 
 is a small system given its relatively short orbital period ($\sim$2.41 hours; 
 \cite{newton_dip}, 2011). 
 It should be emphasized that all three previously known BHCs with short orbital periods 
 below 6 hours (Swift J1753.5--0127 (3.2 hours; \cite{swiftj1753}), 
 XTE J1118+480 (4.1 hours; \cite{xtej1118}), 
 and GRO J0422+32 (5.1 hours; \cite{groj0422}) remained in the hard state 
 during the entire duration of their outbursts. 
 This is probably due to a low mass accretion rate--small 
 accretion disks tend to accumulate less matter than larger disks \citep{shortp}.
 As the hardness-intensity diagram seen in \citet{rxte_timing} and \citet{rxte_spectiming}
  clearly displays a Q-shaped track during the outburst evolution,  
 MAXI J1659--152 might be the first BHC with spectral transitions 
 among the short period BHCs. In addition, all the sources with very short 
 periods are located at high Galactic latitudes (($l$,$b$)=(5.5, +16.5) for MAXI J1659--152, 
 (157.7, +62.3) for XTE J1118+480, (24.9, +12.2) for Swift J1753.5--0127, and (165.9, --11.9) 
 for GRO J0422+32). XTE J1118+480 is thought to be a run-away microquasar which is moving at
 a high speed of about 100 km s$^{-1}$ \citep{xtej1118_2}, and may have been born 
 during a supernova on the Galactic disk, and then subsequently moved toward the 
 Galactic halo by kicks received during the explosion. This might suggest that MAXI J1659--152
 is also a run-away microquasar similar to XTE J1118+480. 
 The long-term monitoring of radio and/or optical observations of the MAXI J1659--152 
 source location with high spatial accuracy will be useful for determining the origin 
 of this source. Also, the recently discovered BH transient Swift
 J1357.2--0933 lies well above the Galactic plane ($l$,$b$)=(328.7, +50.0) and stays in 
 the hard state throughout the outburst \citep{swiftj1357}. 
 Swift J1357.2--0933 may thus be a short-period black hole candidate \citep{swiftj1357_2}.

\subsection{Estimation of black hole mass}\label{discuss2}

 We have evaluated the innermost radius using the
 {\tt simpl} Comptonization model, and have found that it started 
  at about 110 $d_{10}{\cos i}^{-1/2}$km, where $d_{10}$ is the distance in units of 10 kpc and $i$ is the inclination angle,  during the initial phase and gradually approached the constant value of
 $\sim$35 $d_{10}{\cos i}^{-1/2}$km. This could imply that the innermost radius finally reached the ISCO \citep{constrin_isco}. Assuming that the ISCO coincides with three times the 
 Schwarzschild radius (3$R_{\rm S}$), as in the case of a non-spinning Schwarzschild BH, 
 we can estimate the mass of the central object in MAXI J1659--152. However, 
 we note that the observed innermost 
 radius ($r_{\rm in}$) does not reflect a physical radius ($R_{\rm in}$) due to 
 two main effects: the MCD model does not assume the torque-free condition 
 at the innermost radius; and 
 the emergent spectrum is distorted by a hardening due to the electron 
 scattering, resulting in a blackbody component 
 with a higher color temperature ($T_{\rm col}$) than the effective temperature ($T_{\rm eff}$). 
 We follow the correction method for these two effects that has been developed by \citet{rin_cor}. 

 \citet{simpl_appl} applied the {\tt simpl} model to energy spectra taken during 
  intermediate states, and showed that the innermost radius can increase
  when the Compton fraction is larger than 0.25. 
 So, using the averaged $r_{\rm in}$ value of 35.3 km for $i$=0$^{\circ}$ and $D$=10 kpc,
 weighted for observations with a Compton fraction smaller than 0.25, 
 we can estimate the BH mass as follows:

\begin{equation}
R_{\rm in} = \xi \kappa^2 r_{\rm in} = 59.4 \left( \frac{D}{10 {\rm kpc}} \right) \left(\frac{\cos i}{\cos 60^{\circ}} \right) ^{-\frac{1}{2}} {\rm km} = 82.6 \left( \frac{D}{10 {\rm kpc}} \right) \left(\frac{\cos i}{\cos 75^{\circ}} \right) ^{-\frac{1}{2}} {\rm km},
\end{equation} where $\xi$ is 0.412 \citep{rin_cor} and $\kappa=\frac{T_{\rm col}}{T_{\rm eff}}$ is the spectral hardening factor 
 (1.7: \cite{hardening}). Since we assume $R_{\rm in}$ to be 3$R_{\rm S}$ ($R_{\rm S}$=2$GM/c^2$), the 
obtained black hole mass $M$ is 
\begin{equation}\label{bh_mass}
M = \frac{c^2 R_{\rm in}}{6G} = 6.71 \left( \frac{D}{10 {\rm kpc}} \right) \left( \frac{\cos i}{\cos 60^{\circ}} \right) ^{-\frac{1}{2}} M_{\odot} = 9.32 \left( \frac{D}{10 {\rm kpc}} \right) \left( \frac{\cos i}{\cos 75^{\circ}} \right) ^{-\frac{1}{2}} M_{\odot}.
\end{equation}

\noindent In the case of a rapidly rotating Kerr black hole, the ISCO approaches much closer to the black hole
 to a minimum of 0.5$R_{\rm S}$, depending on the spin parameter $a$. 
In the maximum rotating case ($a\approx$1), the BH mass estimate can be larger by 
a maximum factor of 6. 

\begin{figure}
  \begin{center}
    \FigureFile(80mm,80mm){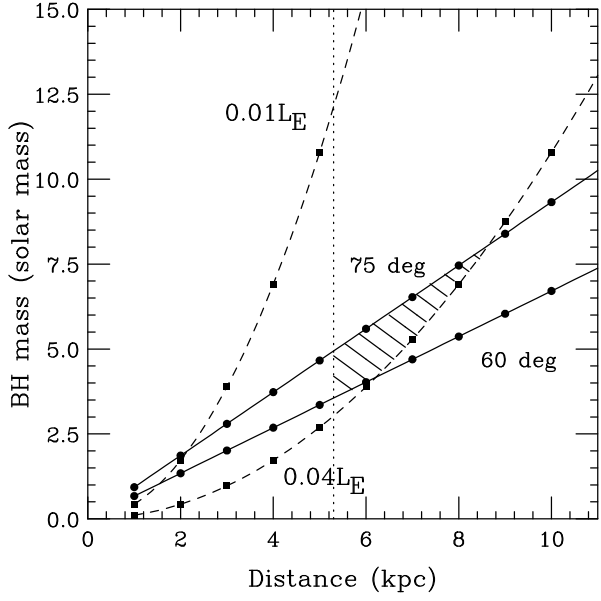}
 \end{center}
 \caption{The allowed range of the black hole mass for MAXI J1659--152, shown by the hatched region. The solid and dashed lines show constraints from the inclination angle of 60--75$^{\circ}$ and the transition luminosity to the hard state (1--4 \% Eddington luminosity), respectively. The dotted line shows a lower limit using the distance 5.3 kpc. }\label{fig11}
\end{figure}

 For the mass constraint, we need information for a distance $D$ and an 
 inclination angle $i$ of MAXI J1659--152. We use estimates ($D>$5.3 kpc and 
 $i$=60--75$^{\circ}$) suggested by \citet{dip} and \citet{swift_dip}. 
 Furthermore, it is known that the thermal-to-hard transition will occur at 1--4 \% of the 
 Eddington luminosity ($L_{\rm E}$ = 1.25$\times 10^{38} \frac{M}{M_{\odot}}$ erg s$^{-1}$) 
 in soft X-ray transients \citep{lum_trans}. The observed 
 flux from the hard state after the transition (MJD 55508, day 44.1) was 1.97$\times$10$^{-9}$ 
 erg cm$^{-2}$ s$^{-1}$ in the 3--50 keV range. This was then converted to a bolometric 
 flux of 4.50$\times$10$^{-9}$ erg cm$^{-2}$ s$^{-1}$, assuming a high energy cutoff 
 at 200 keV. Using these three constraints, we plotted the allowed BH mass range by changing 
 the distance and inclination angle in Figure \ref{fig11}. The BH mass is constrained to 
  3.6--8.0 $M_{\odot}$ in the 5.3--8.6 kpc range. 

 For more accurate estimations of disk parameters, 
 we applied the {\tt kerrbb} model \citep{kerrbb} to the RXTE data instead of 
 the {\tt diskbb} model. The {\tt kerrbb} model takes into account relativistic 
 effects (e.g. light bending around the spinning Kerr BH). The spin parameter $a$ 
 was poorly constrained for all the observations, so we fixed $a$ to a 
 value of 0 (i.e. non-spinning BH). The fits using this model were reasonable (average $\chi^2$/dof 
  = 56.1/73 for $i$=60$^{\circ}$ and 56.2/73 for $i$=75$^{\circ}$). Assuming $D$=10 kpc, 
 the BH mass is estimated at 6.7 $M_{\odot}$ for $i$=60$^{\circ}$ and 9.9 $M_{\odot}$ for $i$=75$^{\circ}$, 
 which is in a good agreement with the estimate obtained from the MCD model (equation \ref{bh_mass}). 
 We also applied the {\tt kerrbb} model to simultaneous Swift and RXTE data in both period G 
  and period H, when the Compton fraction was lower than 0.25. The fitting parameters are shown in 
  Table \ref{tab3}. Assuming $D$=10 kpc for comparison with the MCD results,
  the mass is estimated to be 6.18$\pm$0.21 $M_{\odot}$ (period G) and 
   5.74$\pm$0.21 $M_{\odot}$ (period H) for $i$=60$^{\circ}$, and 9.09$\pm$0.31 $M_{\odot}$ (period G) 
  and 8.46$\pm$0.31 $M_{\odot}$ (period G) for $i$=75$^{\circ}$.  
 The lower limit is 3.27$\pm$0.11 $M_{\odot}$ (period G) 
 and 3.04$\pm$0.11 $M_{\odot}$ (period H) for $i=60^{\circ}$ and $D$=5.3 kpc. These results 
 support the conclusion that the primary object in MAXI J1659--152 is a black hole.

 \citet{rxte_spectiming2} estimated the black hole mass to be 20$\pm$3 $M_{\odot}$ using the 
  empirical correlation among the photon index $\Gamma$, the LFQPO frequency,
  and the normalization in {\tt bmc}. 
 This value is much larger than our estimated one. This discrepancy could be resolved
  by taking into account the BH spin in our estimation, since the BH mass 
  can increase up to 31--42 $M_{\odot}$ for the maximum spinning BH ($a\approx$1) 
 from 5.1--7.1 $M_{\odot}$ for the non-spinning BH ($a$=0), assuming the estimated 
  distance of 7.6 kpc. Fixing the mass at 20$M_{\odot}$, the {\tt kerrbb} model fitting 
   for the period G gives $a$=0.91$\pm$0.01 for $i$=60$^{\circ}$ and 
  $a$=0.84$\pm$0.01 for $i$=75$^{\circ}$.
  Dynamic mass measurements using optical observations are expected give more stringent 
 values, and will resolve this issue in the future.

\begin{table}
\begin{center}
{\footnotesize
\caption{Best-fit parameters with {\tt kerrbb} model.}\label{tab3}
\begin{tabular}{cccccc}\\\hline\hline

Model & Par. & \multicolumn{2}{c}{Period G (Day 24.0)} & \multicolumn{2}{c}{Period H (Day 27.0)}\\
      &      &  $i$=60$^{\circ}$  & $i$=75$^{\circ}$  & $i$=60$^{\circ}$ & $i$=75$^{\circ}$    \\\hline
wabs   & $N_{\rm H}$\footnotemark[$*$]  & 0.23$^{+0.01}_{-0.01}$ & 0.22$^{+0.01}_{-0.01}$ & 0.27$^{+0.01}_{-0.01}$ & 0.26$^{+0.01}_{-0.01}$ \\ 
kerrbb\footnotemark[$\dagger$]  & $M_{\rm BH}$               & 6.18$^{+0.21}_{-0.21}$ & 9.09$^{+0.31}_{-0.30}$ & 5.74$^{+0.21}_{-0.20}$ & 8.46$^{+0.31}_{-0.30}$ \\ 
($D$=10 kpc)& $\dot{M}$\footnotemark[$\ddagger$]              & 2.32$^{+0.03}_{-0.03}$ & 3.81$^{+0.06}_{-0.06}$ & 2.11$^{+0.03}_{-0.03}$ & 3.47$^{+0.05}_{-0.05}$ \\
kerrbb\footnotemark[$\dagger$]  & $M_{\rm BH}$               & 3.27$^{+0.11}_{-0.11}$ & 4.82$^{+0.17}_{-0.16}$ & 3.04$^{+0.11}_{-0.11}$ & 4.48$^{+0.16}_{-0.16}$ \\ 
($D$=5.3 kpc)  & $\dot{M}$\footnotemark[$\ddagger$]           & 0.65$^{+0.01}_{-0.01}$ & 1.07$^{+0.02}_{-0.02}$ & 0.59$^{+0.01}_{-0.01}$ & 0.97$^{+0.01}_{-0.01}$ \\
simpl  & $\Gamma$                   & 2.34$^{+0.04}_{-0.04}$ & 2.34$^{+0.04}_{-0.04}$ & 2.30$^{+0.04}_{-0.04}$ & 2.30$^{+0.04}_{-0.04}$ \\
             & $f$                  & 0.14$^{+0.01}_{-0.01}$ & 0.14$^{+0.01}_{-0.01}$ & 0.20$^{+0.01}_{-0.01}$ & 0.20$^{+0.01}_{-0.01}$ \\
highecut  & $E_{\rm f}$(keV)        & ---                    & ---                    & ---                    & ---    \\
smedge & $\tau$                     & 0.20$^{+0.27}_{-0.20}$ & 0.23$^{+0.27}_{-0.23}$ & 0.35$^{+0.27}_{-0.28}$ & 0.37$^{+0.27}_{-0.27}$\\ 
constant & factor                   & 0.93$^{+0.01}_{-0.01}$ & 0.93$^{+0.01}_{-0.01}$ & 0.87$^{+0.01}_{-0.01}$ & 0.87$^{+0.01}_{-0.01}$\\
$\chi^2$/d.o.f. &                   & 560.6/462              & 561.8/462	       & 645.3/495  & 645.6/495                 \\\hline 
\multicolumn{4}{@{}l@{}}{\hbox to 0pt{\parbox{100mm}{\footnotesize 
\hspace{1mm}
\par\noindent
\footnotemark[$*$] In unit of 10$^{22}$ cm$^{-2}$.
\par\noindent
\footnotemark[$\dagger$] Either one of two cases are applied.  
\par\noindent
\footnotemark[$\ddagger$] A mass accretion rate in unit of 10$^{18}$ g s$^{-1}$.
}\hss}}
\end{tabular}
}
\end{center}
\end{table}


\section{Summary}

We have presented timing and spectral results from 65 RXTE pointed observations 
 and 8 simultaneous Swift and RXTE observations of 
the new  X-ray transient MAXI J1659--152 in the 2010 outburst. 
The main results of the study are:  

\begin{enumerate}

\item The outburst lasted about two months, which is 
  shorter than typical outbust duration of BHCs (more than 100 days). 
 Most of the observation epochs are classified into the intermediate state based on definitions 
 in \citet{review1}. 
 MAXI J1659-152 might be the first among BHCs with 
 orbital periods shorter than 6 hours to display spectral transitions.

\item X-ray spectral and timing parameters of MAXI J1659--152 are similar 
 to other black hole candidates. The energy spectra in both the 0.6--50 keV and 3--50 keV ranges
  are well modeled by the MCD model plus a power-law with an exponential cutoff component
  modified with the smeared edge and Galactic absorption. 
 Both type-B and type-C QPOs were found in the power spectra among 51 observations, 
 when both the disk and corona components are securely present. 

\item The maximum innermost temperature was $\sim$0.9 keV. This is low compared to typical values of
   $T_{\rm in}$ observed in bright BHCs (1.0--1.3 keV). The high energy cutoff was 
  present at 30--40 keV during the initial phase of the outburst. This may be due to the 
 Compton cooling of the corona by soft photons at the beginning of the outburst. 

\item The central frequency for the type-C QPOs, seen in the range of 1.6--7.3 Hz, is correlated with 
 the innermost temperature, disk flux, and photon index, and is anti-correlated with rms variability 
 and innermost radius. This can be explained by the scenario that the type-C QPO originates from the 
 truncated disk radius. The type-B QPO frequency varied from 1.7 to 4.1 Hz while 
 the innermost radius remained constant at 35 km. Hence, the origin of the type-B QPO is 
 definitely different from that of the type-C QPO. 
    
\item Using combined Swift and RXTE data, we found that the innermost radius decreased by 
  a factor of more than 3 and approached a constant value of 35 km for assumed distance of 10 kpc 
 and inclination of 0 degree during the middle of the outburst. 
 Assuming this constant value to be the ISCO and the black hole to be non-spinning, the
 black hole mass of MAXI J1659--152 is estimated at 3.6--8.0 $M_{\odot}$,
 for a distance of 5.3--8.6 kpc and an inclination angle of 60--75$^{\circ}$ 
 (\cite{dip}, \cite{swift_dip}). 

\end{enumerate}

We thank the anonymous referee for his/her useful comments and discussions.
This research has made use of: MAXI data provided by RIKEN, JAXA and the MAXI team; 
 Swift/BAT transient monitor results provided by the Swift/BAT team; RXTE/ASM 
 data provided by the ASM/RXTE teams at MIT and at the RXTE SOF and GOF at NASA/GSFC; RXTE/PCA data obtained from the High Energy Astrophysics Science Archive 
Research Center (HEASARC) provided by NASA/GSFC; and 
 Swift/XRT data supplied by the UK Swift Science Data Centre at the University of Leicester.




\end{document}